\documentstyle[preprint,aps,graphicx,floats]{revtex}

\begin{document}

\preprint{$
\begin{array}{l}
\mbox{CERN-TH/2002-327}\\[-3mm]
\mbox{UB-HET-02-05}\\[-3mm]
\mbox{FERMILAB-Pub-02/199-T}\\[-3mm]
\mbox{DESY-02-183}\\[-3.mm]
\mbox{November~2002} \\ [3mm]
\end{array}
$}

\title{
Determining the Higgs Boson Self Coupling at Hadron Colliders}

\author{U.~Baur\footnote{e-mail: baur@ubhex.physics.buffalo.edu\\[-12.mm]}}
\address{Department of Physics,
State University of New York, Buffalo, NY 14260, USA\\[-2.mm]}
\author{T.~Plehn\footnote{e-mail: tilman.plehn@cern.ch\\[-12.mm]}}
\address{CERN Theory Group, CH-1211 Geneva 23, Switzerland\\[-2.mm] }
\author{D.~Rainwater\footnote{e-mail: rain@mail.desy.de}}
\address{DESY Theorie, Notkestrasse 85, D-22603 Hamburg,
Germany\\[-3.mm]
and\\[-3.mm]
Theory Group, Fermi National Accelerator Laboratory, Batavia, IL 60510, USA} 

\maketitle 

\begin{abstract}
\baselineskip13.pt  
  Inclusive Standard Model Higgs boson pair production at hadron colliders
has the capability to determine the Higgs
  boson self-coupling, $\lambda$. We present a detailed analysis of the
$gg\to HH\to (W^+W^-)(W^+W^-)\to (jj\ell^\pm\nu)(jj{\ell'}^\pm\nu)$ and 
$gg\to HH\to (W^+W^-)(W^+W^-)\to (jj\ell^\pm\nu)({\ell'}^\pm\nu
{\ell''}^\mp\nu)$ ($\ell,\,{\ell'},\,{\ell''}=e,\,\mu$) 
signal channels, and the relevant background processes, for the
  CERN Large Hadron Collider, and a future Very Large Hadron Collider
operating at a center-of-mass energy of 200~TeV. We also derive
quantitative sensitivity 
limits for $\lambda$. We find that it
  should be possible at the LHC with design luminosity to establish
  that the Standard Model Higgs boson has a non-zero self-coupling and
  that $\lambda / \lambda_{SM}$ can be restricted to a range of 0~--~3.8
  at $95\%$ confidence level (CL) if its mass is between 150 and
200~GeV. At a 200~TeV collider with an integrated luminosity of
300~fb$^{-1}$, $\lambda$ can be determined with an accuracy of $8 - 25\%$ at
$95\%$ CL in the same mass range. 
\end{abstract}

\newpage


\tightenlines

\section{Introduction}

The CERN Large Hadron Collider (LHC) is widely regarded as capable of
directly observing the agent responsible for electroweak symmetry
breaking and fermion mass generation. This is generally believed to be
a light Higgs boson with mass $m_H<200$~GeV~\cite{lepewwg}, a mass
region for which discovery is covered by multiple channels, most
notably by the decay that will be of interest here, $H\to
W^+W^-$~\cite{H2WW}. Once a Higgs boson candidate has been observed,
emphasis will shift to a precise determination of its properties. The
LHC promises complete coverage of Higgs decay scenarios~\cite{tdr+},
including general parameterizations in the Minimal Supersymmetric
Standard Model (MSSM)~\cite{tdr+,wbf_ll}, invisible Higgs
decays~\cite{wbf_inv}, and possibly even Higgs boson decays to
muons~\cite{Hmumu}.  With mild theoretical assumptions and an
integrated luminosity of 200~fb$^{-1}$, the Higgs boson total width,
$\Gamma_H$, and the gauge and various Yukawa couplings can be
determined~\cite{Hcoup,Yt,mrw} with a precision of
$10-30\%$~\cite{zep}. At an $e^+e^-$ linear collider with a center of
mass energy of 350~GeV or more, these measurements can be improved by
up to a factor~10~\cite{LC}, if an integrated luminosity of
500~fb$^{-1}$ can be achieved.

While these studies have shown that future colliders promise broad and
significant capability to measure various properties of the Higgs
sector, what remains is to determine the actual Higgs potential. This
appears in the Lagrangian as
\begin{equation}
\label{eq:Hpot}
V(\Phi) \, = \, 
-\lambda v^2 (\Phi^\dagger\Phi) \, + \, \lambda (\Phi^\dagger\Phi)^2,
\end{equation}
where $\Phi$ is the Higgs field, $v=(\sqrt{2}G_F)^{-1/2}$ is the
vacuum expectation value, and $G_F$ is the Fermi constant. In the
Standard Model (SM), 
\begin{equation}
\label{eq:lamsm}
\lambda=\lambda_{SM}={m_H^2\over 2v^2}\,.
\end{equation}
Regarding the SM as an effective theory, the Higgs boson self-coupling
$\lambda$ is {\it per se} a free parameter. $S$-matrix unitarity
constrains $\lambda$ to $\lambda\leq 8\pi/3$~\cite{unit}. An anomalous
Higgs boson self-coupling appears in various beyond the SM scenarios,
such as models with a composite Higgs boson~\cite{georgi}, or in two
Higgs doublet models, for example the MSSM~\cite{mssm}. To measure
$\lambda$, and thus determine the Higgs potential, at a minimum
experiments must observe Higgs boson pair production.

Strictly speaking, both the trilinear Higgs boson coupling $g_{HHH}$
and the quartic coupling $g_{HHHH}$ have to measured separately in
order to fully determine the Higgs potential. While $g_{HHH}$ can be
measured in Higgs pair production, triple Higgs production is needed
to probe $g_{HHHH}$. Since the cross sections for $HHH$ production
processes are more than a factor $10^3$ smaller than those for Higgs
pair production at linear colliders~\cite{LC_HH3}, and about an order
of magnitude smaller at hadron colliders~\cite{higgs_self}, the
quartic Higgs boson coupling will likely remain elusive even at the
highest collider energies and luminosities considered so far. In the
following we therefore restrict ourselves to $g_{HHH}$ which is
related to $\lambda$ by $g_{HHH}=3\lambda v$.

Several studies of Higgs pair production in $e^+e^-$ collisions have
been conducted over the past few
years~\cite{LC_HH3,LC_HH1,LC_HH1a,LC_HH2}, and quantitative
sensitivity limits for $\lambda$ have been derived for several
proposed linear colliders with center of mass energies spanning the
range from $\sqrt{s}=500$~GeV to 3~TeV. For example, a study employing
neural net techniques found that $\lambda$ can be measured with a
precision of about $20\%$ at a linear collider with $\sqrt{s}=500$~GeV and
an integrated luminosity of 1~ab$^{-1}$, if
$m_H=120$~GeV~\cite{LC_HH1a}. In contrast, the potential of the LHC to
probe the Higgs boson self coupling has begun to be explored only
recently. A survey of Higgs pair production and background processes
at an upgraded LHC, which would gather 20~times the amount of data
expected in the first run (dubbed SLHC), was presented in
Ref.~\cite{SLHC}. In Ref.~\cite{BPR}, we discussed the prospects for
determining $\lambda$ at the LHC with design luminosity in Higgs pair
production via gluon fusion and subsequent decay to same-sign
dileptons via weak gauge $W^\pm$ bosons,
\begin{equation}
\label{eq:samesign}
gg\to HH\to (W^+W^-)(W^+W^-)\to (jj\ell^\pm\nu)(jj{\ell'}^\pm\nu),
\end{equation}
where $\ell$ and ${\ell'}$ are any combination of electrons and muons,
and presented quantitative estimates of sensitivity limits for
$\lambda$ for $150~{\rm GeV}\leq m_H\leq 200$~GeV and various
integrated luminosities. Finally, in Ref.~\cite{blondel}, a {\sc
  pythia}~\cite{pythia} based study of several final states resulting
from $gg\to HH\to (W^+W^-)(W^+W^-)$ and $gg\to HH\to (W^+W^-)(ZZ)$ for
LHC and SLHC was conducted.

In this paper we present a more detailed and extended analysis of
Higgs pair production via gluon fusion at hadron colliders. In
Ref.~\cite{BPR}, we included only the two largest sources of
background, $W^\pm W^+W^-jj$ and $t\bar{t}W^\pm$ production, in our
calculation. The effect of the remaining background contributions on
the sensitivity limits for $\lambda$ was estimated by scaling the
combined $WWWjj$ and $t\bar{t}W$ cross section by a factor~1.1, as
suggested by Ref.~\cite{SLHC}, which found that the remaining
backgrounds are small. Here we present a more complete calculation of
the background which includes, in addition to $WWWjj$ and $t\bar{t}W$
production, $W^\pm W^\pm jjjj$, $W^\pm Zjjjj$, $t\bar{t}Z$,
$t\bar{t}j$, $t\bar{t}t\bar{t}$, $W^+W^-W^+W^-$ and $WWZjj$
production.  Furthermore, we discuss the potential size of backgrounds
arising from overlapping events and double-parton scattering.  We also
extend our previous analysis by considering the three lepton channel
\begin{equation}
\label{eq:lll}
gg\to HH\to (W^+W^-)(W^+W^-)\to (jj\ell^\pm\nu)({\ell'}^\pm\nu
{\ell''}^\mp\nu)
\end{equation}
($\ell,\,{\ell'},\,{\ell''}=e,\,\mu$), and the background processes
which affect it. In addition to the LHC and SLHC, we calculate signal
and background cross sections, and derive sensitivity bounds for
$\lambda$, for a Very Large Hadron Collider (VLHC), assuming a $pp$
collider operating at $\sqrt{s}=200$~TeV with a luminosity of ${\cal
  L}=2\times 10^{34}~{\rm cm^{-2}\, s^{-1}}$. These parameters
correspond to one of the options listed in Ref.~\cite{vlhc}.

The remainder of this paper is organized as follows. In
Secs.~\ref{sec:sec2} and~\ref{sec:sec3} we outline our calculation of
signal and background processes for the same-sign dilepton and three
lepton final states, respectively. In Sec.~\ref{sec:sec4}, we derive
sensitivity limits for $\lambda$ for various integrated luminosities
at the LHC and VLHC. Our conclusions are given in Sec.~\ref{sec:sec5}.

\section{The same-sign dilepton final state}
\label{sec:sec2}

There are several mechanisms for pair production of Higgs bosons in
hadronic collisions. Higgs boson pairs can be produced via gluon
fusion, $gg\to HH$~\cite{higgs_self,lhc_hh}, weak boson fusion, $qq\to
qqHH$~\cite{wbf}, associated production with $W$ or $Z$ bosons, $q\bar
q\to VHH$, $V=W,\,Z$~\cite{assoc}, and associated production with
$t\bar{t}$ pairs, $gg,\,q\bar q\to t\bar{t}HH$~\cite{SLHC}. At the LHC,
inclusive Higgs boson pair production is dominated by gluon fusion.
The weak boson fusion process, and associated production with $W/Z$
bosons or $t\bar{t}$ pairs yield cross sections which are about a
factor~10 and~30 smaller than that for $gg\to HH$~\cite{lhc_hh,SLHC}.
Since Higgs pair production at the LHC is rate limited, we concentrate
on the gluon fusion process in the following.

For $m_H<140$~GeV, the dominant decay mode of the SM Higgs boson is
$H\to b\bar{b}$, and the QCD $b\bar{b}b\bar{b}$ background overwhelms
the $gg\to HH$ signal~\cite{4b}. For $m_H>140$~GeV, $H\to W^+W^-$
dominates, and the $W^+W^-W^+W^-$ final state has the largest
individual branching ratio. If all $W$ bosons decay hadronically, QCD
multi-jet production dwarfs the signal. A similar result is obtained
for the $\ell^\pm\nu+6$~jet (only one $W$ boson decays leptonically),
and $\ell^\pm\nu{\ell'}^\mp\nu+4$~jet (one $W^+W^-$ pair decays
leptonically) final states, where $W+$~multi-jet and
$W^+W^-+$~multi-jet production provide very large backgrounds. This
leaves the same-sign dilepton final states,
$(jj\ell^\pm\nu)(jj{\ell'}^\pm\nu)$, modes where three $W$ bosons
decay leptonically and one decays hadronically, and the all-leptonic
decay modes. The latter suffer from a large suppression due to the
small $WWWW\to 4\ell+4\nu$ branching ratio of $(0.216)^4=0.0022$
(BR($W\to\ell\nu)=0.216$, $\ell=e,\,\mu$). In the following we
therefore only consider the $(jj\ell^\pm\nu)(jj{\ell'}^\pm\nu)$ and
$(jj\ell^\pm\nu)({\ell'}^\pm\nu {\ell''}^\mp\nu)$ final states.

In this section we discuss in detail the calculation of signal and
background cross sections for the $(jj\ell^\pm\nu)(jj{\ell'}^\pm\nu)$
final state. The three lepton final state will be considered in
Sec.~\ref{sec:sec3}.

\subsection{Calculation of the signal cross section}

\begin{figure}[t] 
\begin{center}
\includegraphics[width=15cm]{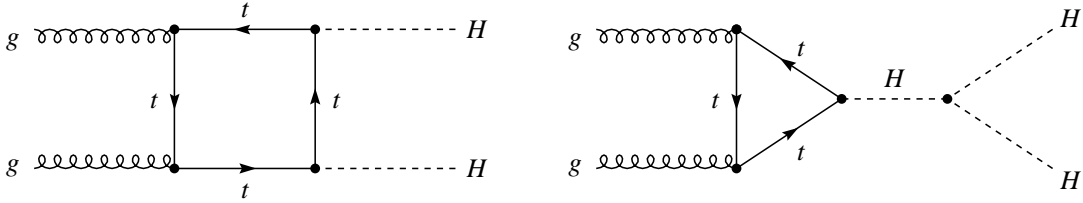}
\vspace*{2mm}
\caption[]{\label{fig:fig1} 
Representative Feynman diagrams for the process $gg\to HH$. }
\end{center}
\end{figure}

The Feynman diagrams contributing to $gg\to HH$ in the SM consist of
fermion triangle and box diagrams (see
Fig.~\ref{fig:fig1})~\cite{higgs_self}.
Non-standard Higgs boson self-couplings only affect the triangle
diagrams with a Higgs boson exchanged in the $s$-channel. We calculate
the $gg\to HH\to (W^+W^-)(W^+W^-)\to
(jj\ell^\pm\nu)(jj{\ell'}^\pm\nu)$ cross section using exact loop
matrix elements~\cite{higgs_self}. As demonstrated in Ref.~\cite{BPR},
the infinite top quark mass limit, which is commonly used in place of
exact matrix elements to speed up the calculation, reproduces the
correct total cross section for $HH$ production to within $10\%$ to
$30\%$ for Higgs masses between 140~GeV and 200~GeV, but produces
completely incorrect kinematic distributions. The intermediate Higgs
and $W$ bosons are treated off-shell using finite widths in the double
pole approximation in our calculation. Decay correlations for the
$H\to W^+W^-\to 4$~fermion decays are fully taken into
account~\cite{glover}.

Signal results are computed consistently to leading order QCD with the
top quark mass set to $m_t=175$~GeV and SM $HWW$ and top quark Yukawa
couplings, and the renormalization and factorization scales are taken
to be the Higgs boson mass~\cite{higgs_self}. The contributions of all
other quarks to the $gg\to HH$ box and triangular diagrams are
strongly suppressed due to their $Hf\bar f$ couplings, which are
proportional to the fermion mass.  The effects of next-to-leading
(NLO) order QCD corrections are included in our calculation by
multiplying the differential cross section by an overall factor
$K=1.65$ at LHC and $K=1.35$ at VLHC energies ($K$-factor) for scale
choice $\mu = m_H$, as suggested by Ref.~\cite{higgs_nlo} where the
QCD corrections for $gg\to HH$ have been computed in the large $m_t$
limit. Although this approximation cannot replace a calculation of the
full NLO QCD corrections to $gg\to HH$, it is expected to work well in
this particular case: it is well known from single Higgs boson
production via gluon fusion~\cite{spira} that the two--loop QCD
corrections to the one--loop Higgs production amplitude are well
approximated by multiplying the leading order one--loop cross section
for a finite top quark mass by the $K$-factor obtained in the large
$m_t$ limit.  This feature can be easily understood by recalling that
the dominant corrections originate from radiation off the initial
state gluons, which is universal.

The lowest order $gg\to HH$ cross section exhibits a rather strong
dependence on the renormalization and factorization scales. For
example, for $m_H=200$~GeV and $\mu^2 = m_H^2$, and using exact loop
matrix elements, one obtains a total cross section of 8.26~fb. For
$\mu^2=\hat{s}$, on the other hand, one finds a cross section which is
almost a factor~1.5 smaller. NLO QCD corrections, albeit computed in
the large $m_t$ limit, significantly reduce the $\mu$ dependence:
varying the scale $\mu$ from $\mu^2 = m_H^2$ to $\mu^2=\hat{s}$, the
cross section including the $K$-factor decreases by a factor~1.25
instead of~1.5 without the $K$-factor.

For $\mu^2 = m_H^2$, the $K$-factor at VLHC energies is smaller than
that obtained for the LHC. At higher energies, smaller parton momentum
fractions are probed. This results in an increased sensitivity of the
cross section to the choice of factorization scale which partially
compensates the variation of the cross section with the
renormalization scale.

In all our calculations we use a value for the strong coupling
constant of $\alpha_s(M_Z) = 0.1185$~\cite{long}. All signal and
background cross sections are computed using CTEQ4L~\cite{cteq} parton
distribution functions.

The kinematic acceptance cuts for both signal and backgrounds in the
$(jj\ell^\pm\nu)(jj{\ell'}^\pm\nu)$ channel at the LHC and VLHC are:
\begin{eqnarray}
\label{eq:cuts1}
&p_T(j) > 30,\, 30,\, 20,\, 20~{\rm GeV} , \qquad 
p_T(\ell) > 15,\,15~{\rm GeV}    ,        \\
&|\eta(j)| < 3.0     ,        \qquad \qquad \qquad \qquad 
|\eta(\ell)| < 2.5   ,              \\
&\Delta R(jj) > 0.6 ,   \qquad 
\Delta R(j \ell) > 0.4 , \qquad 
\Delta R(\ell \ell) > 0.2 ,
\label{eq:cuts2}
\end{eqnarray}
where $\Delta R =
\left[\left(\Delta\phi\right)^2+\left(\Delta\eta\right)^2\right]^{1/2}$
is the separation in the pseudorapidity -- azimuthal angle plane.  In
addition we require the four jets to combine into two pseudo-$W$ pairs
with invariant masses
\begin{equation}
\label{eq:jjmass}
50~{\rm GeV} < m(jj) < 110~{\rm GeV},
\end{equation}
and assume that this captures $100\%$ of the signal and backgrounds.
We do not impose a missing transverse momentum cut which would remove
a considerable fraction of the signal events. Detector resolution
effects are not taken into account in our calculation.

As we have shown in Ref.~\cite{BPR}, at the LHC, the main background
from $WWWjj$ and $t\bar{t}W$ production can be reduced by about
$45\%$, with little impact on the signal, by imposing a more
restrictive jet-jet separation cut of $\Delta R(jj) >1.0$. In
contrast, at VLHC energies, there is little gain in tightening the
$\Delta R(jj)$ cut. Subsequently, we therefore require
\begin{eqnarray}
\Delta R(jj) >1.0 & \qquad \qquad {\rm at~the~LHC,~and}\\
\Delta R(jj) >0.6 & \qquad \qquad {\rm at~the~VLHC}
\end{eqnarray}
in the $(jj\ell^\pm\nu)(jj{\ell'}^\pm\nu)$ channel.

\begin{figure}[t!] 
\begin{center}
\includegraphics[width=13.5cm]{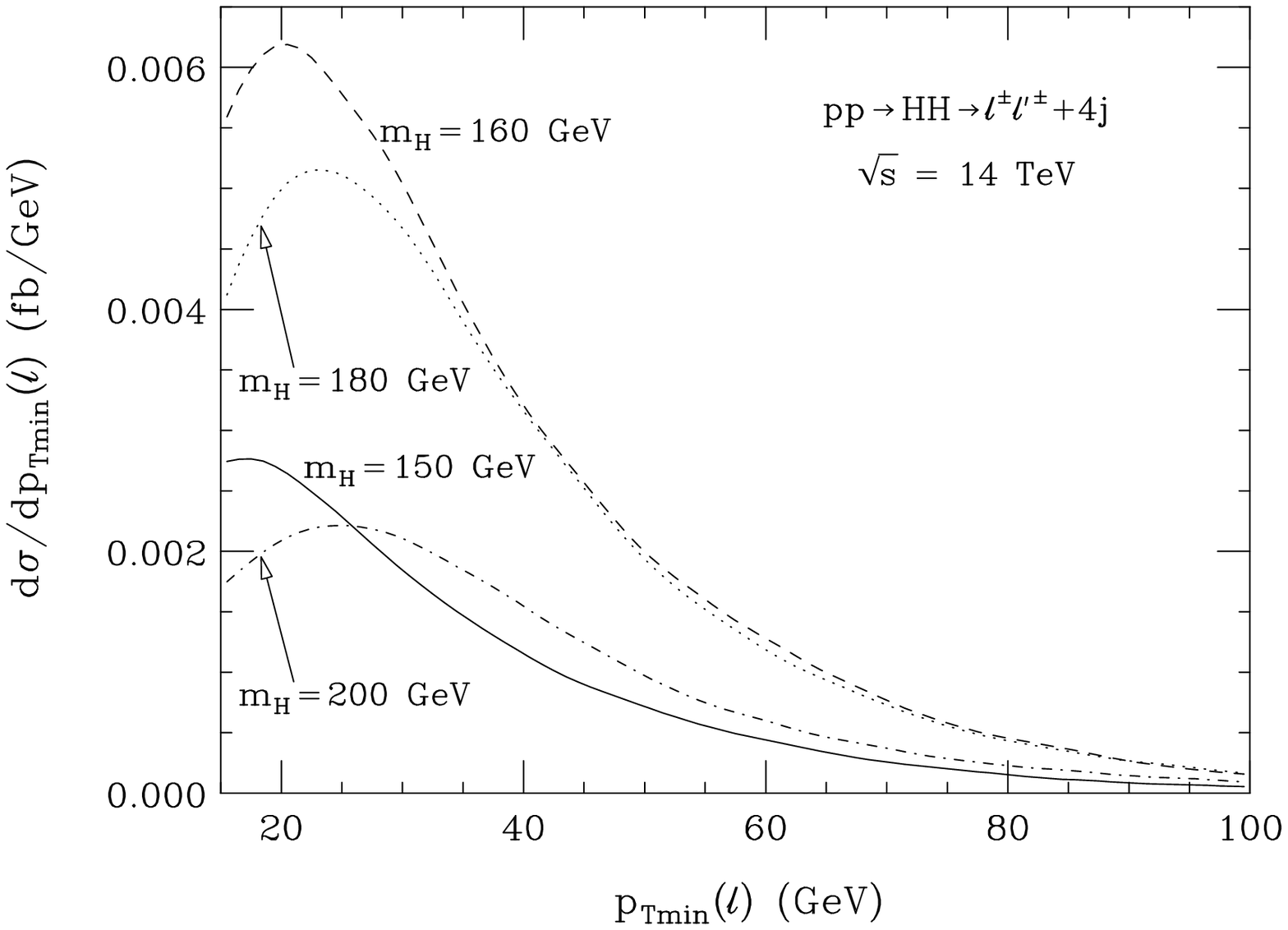}
\vspace*{2mm}
\caption[]{\label{fig:fig2} 
  The $pp\to HH\to (W^+W^-)(W^+W^-)\to \ell^\pm{\ell'}^\pm+4j$
  differential cross section as a function of the minimum lepton
  transverse momentum, $p_{Tmin}(\ell)$, for $pp$ collisions at
  $\sqrt{s}=14$~TeV. Results are shown for $m_H=150$~GeV (solid line),
  $m_H=160$~GeV (dashed line), $m_H=180$~GeV (dotted line), and
  $m_H=200$~GeV (dash-dotted line).}
\end{center}
\end{figure}

\begin{figure}[h!] 
\begin{center}
\includegraphics[width=13.5cm]{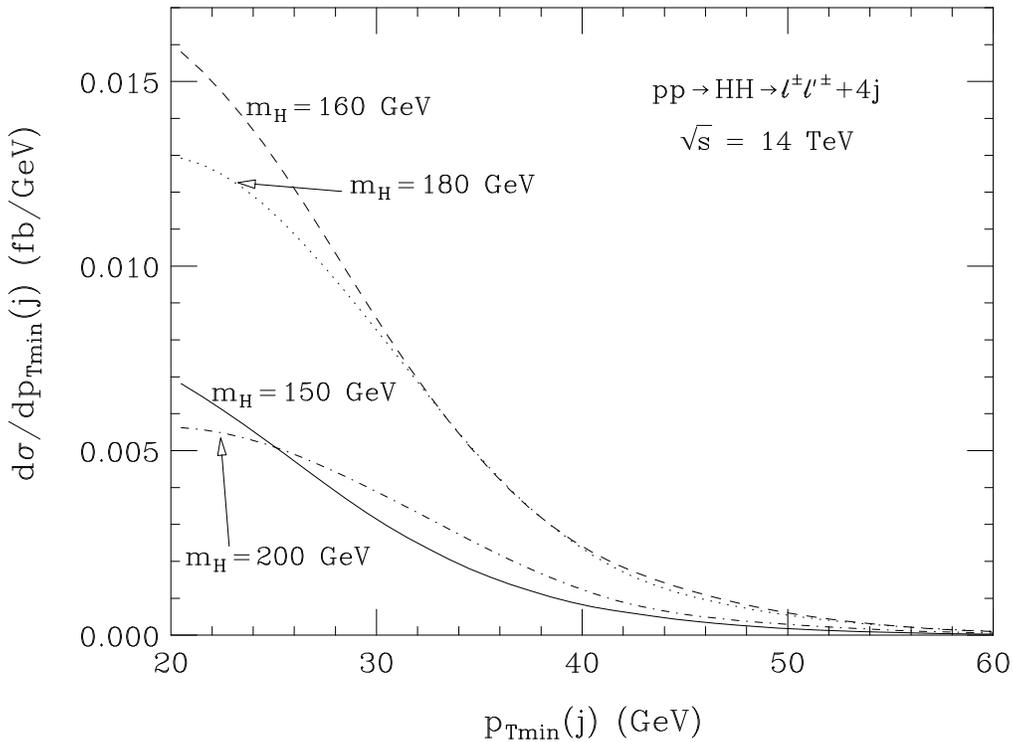}
\vspace*{2mm}
\caption[]{\label{fig:fig3} 
  The $pp\to HH\to (W^+W^-)(W^+W^-)\to \ell^\pm{\ell'}^\pm+4j$
  differential cross section as a function of the minimum jet
  transverse momentum, $p_{Tmin}(j)$, for $pp$ collisions at
  $\sqrt{s}=14$~TeV. Results are shown for $m_H=150$~GeV (solid line),
  $m_H=160$~GeV (dashed line), $m_H=180$~GeV (dotted line), and
  $m_H=200$~GeV (dash-dotted line).}
\end{center}
\end{figure}

Our choice of $p_T$ cuts for jets and leptons in Eqs.~(\ref{eq:cuts1})
--~(\ref{eq:cuts2}) is driven by the goal of retaining as much signal
as possible while ensuring that the LHC experiments, ATLAS and CMS,
can record $\ell^\pm{\ell'}^\pm+4j$ events when operating at the LHC
design luminosity of ${\cal L}=10^{34}~{\rm cm^{-2}\,s^{-1}}$.
Figures~\ref{fig:fig2} and~\ref{fig:fig3} show the $pp\to HH\to
(W^+W^-)(W^+W^-)\to\ell^\pm{\ell'}^\pm+4j$ differential cross section
at the LHC as a function of the lepton and jet minimum transverse
momentum, respectively.
Qualitatively similar results are obtained at VLHC energies. The
differential cross section peaks at low values of $p_{Tmin}$ and falls
very quickly with increasing values of the minimum transverse
momentum, in particular in the jet case. In order to maximize the
signal cross section, the lepton and jet $p_T$ thresholds are thus
chosen as low as possible and yet be compatible with the requirements
of ATLAS and CMS to successfully record such events.

We shall also use the cuts listed in Eqs.~(\ref{eq:cuts1})
--~(\ref{eq:cuts2}) for a luminosity upgraded LHC operating at ${\cal
  L}=10^{35}~{\rm cm^{-2}\,s^{-1}}$, and at the VLHC. Preliminary
studies have concluded~\cite{SLHC,snowm} that cuts similar to those
listed in Eq.~(\ref{eq:cuts1}) --~(\ref{eq:cuts2}) should be
sufficient, although increased background from event pileup is
expected to degrade detector performance, in particular at the SLHC.

\subsection{Calculation of the backgrounds}
\label{sec:bgd}

The SM backgrounds of interest are those that produce two same-sign
leptons and four well-separated jets which reconstruct in two pairs to
a window around the $W$ boson mass. The largest contribution
originates from $W^{\pm}W^+W^-jj$ production, followed by
$t\bar{t}W^\pm$ where one top quark decays leptonically, the other
hadronically, and neither $b$ quark jet is tagged. Other backgrounds
which contribute are: $W^\pm W^\pm jjjj$ production;
$t\bar{t}t\bar{t}$ production, where none of the $b$ quark jets are
tagged, and additional jets or leptons are not observed; $W^\pm Z
jjjj$, $t\bar{t}Z$ and $W^+W^-Zjj$ production with leptonic $Z$ decay
(including off-shell photon interference) where one lepton is not
observed; and $t\bar{t}j$ events where one $b$ quark decays
semileptonically with good hadronic isolation, and the other is not
tagged. In addition, in a high luminosity environment, one has to
worry about backgrounds from overlapping events and double parton
scattering. $b$-quarks are assumed to be tagged with an efficiency of
$50\%$ throughout.  We do not apply an explicit $K$-factor for the
backgrounds here; however, we do later include the potential effect of
QCD corrections on the backgrounds when we extract limits on the Higgs
self-coupling (see Sec.~\ref{sec:sec4}).

\subsubsection{The $WWWjj$, $t\bar{t}W$, $W^+W^-W^+W^-$ and
$t\bar{t}t\bar{t}$ backgrounds} 

We simulate these backgrounds at the parton level using exact matrix
elements generated with {\sc madgraph}~\cite{madgraph}. The $WWWjj$
background has a significant contribution from $WH(\to W^+W^-)jj$
production. For $WWWjj$ production we evaluate the strong coupling
constant $\alpha_s$ and the parton distribution functions at a scale
$\mu$ given by $\mu^2=\sum{p_T^2}$, where the sum extends over all
final state particles; for $t\bar{t}W$ production we take
$\mu=m_t+M_W/2$, and in the $t\bar{t}t\bar{t}$ case we use $\mu=2m_t$.
Top quarks are generated on shell (narrow width approximation), while
all $W$ bosons are allowed to be off shell. Events with one or more
tagged $b$ quarks are rejected. In the $t\bar{t}t\bar{t}$ case we
merge jets if their separation in the pseudorapidity -- azimuthal
angle plane is $\Delta R(jj)<0.6$. The $W^+W^-W^+W^-$ background has a
significant contribution from $WWH(\to WW)$ production. The
$W^+W^-W^+W^-$ cross section at the LHC (VLHC) is found to be a
factor~5 to~25 (25 to~80) smaller than that of $WWWjj$ production,
depending on the mass of the Higgs boson. In this analysis we
therefore ignore the $W^+W^-W^+W^-$ background.

\subsubsection{The $W^\pm W^\pm jjjj$ background}

Although {\sc madgraph} is able to generate exact matrix elements for
$W^\pm W^\pm jjjj$ production, the large number of contributing
Feynman diagrams (more than 6,000) makes a full matrix element based
calculation of the $W^\pm W^\pm jjjj$ background impractical. In order
to estimate the $W^\pm W^\pm jjjj$ cross section we thus have
interfaced the matrix elements for $pp\to W^\pm W^\pm
jj$~\cite{vernon} with {\sc pythia}, which produces the two additional
jets in a leading-log shower approximation. The strong coupling
constant $\alpha_s$ and the parton distribution functions are
evaluated at a scale $\mu$ given by $\mu^2=\sum{p_T^2}$, where the sum
extends over all final state particles.

\subsubsection{The $t\bar{t}Z$, $W^\pm Zjjjj$ and $WWZjj$ backgrounds}

We calculate the $t\bar{t}Z$ cross section using exact matrix
elements. Since more than 15,000 Feynman diagrams contribute to $W^\pm
Zjjjj$ production, we estimate the cross section for this process by
interfacing the $W^\pm Zjj$ matrix elements~\cite{vernon1} with {\sc
  pythia}, similar to $W^\pm W^\pm jjjj$ production. Off-shell photon
interference effects are taken into account in both cases. Both
processes contribute to the background only if one of the leptons from
$Z$/off-shell photon decay is missed. We consider a lepton to be
missed if it has $p_T<10$~GeV or $|\eta|>2.5$. If the lepton is within
a cone of $\Delta R<0.2$ from a detected lepton and has $1~{\rm
  GeV}<p_T<10$~GeV, then the detected lepton is not considered
isolated and the event is rejected. The strong coupling constant and
the parton distribution functions are evaluated in $t\bar{t}Z$ ($W^\pm
Zjjjj$) production at $\mu=m_t+M_Z/2$ ($\mu^2=\sum{p_T^2}$). In order
to avoid the collinear singularity when the missed lepton is collinear
with an observed lepton (which is only relevant if the missed lepton
has $p_T<1$~GeV), finite lepton masses must be maintained in the
calculation of the $W^\pm Zjjjj$ and $t\bar{t}Z$ processes~\cite{mrw},
the latter also including additional considerations to maintain gauge
invariance of the calculation, due to finite top quark width which
must be included for these events.  $t\bar{t}Z$ events are rejected
with a factor 4, which approximates the fraction of events with one or
more tagged $b$ quarks.

The size of the $WWZjj$ background can be estimated from the ratio of
the $WWZ$ and $WWW$ cross sections, together with the suppression
factor which arises from requiring that one lepton is missed, and the
$WWWjj$ rate. We find that the $WWZjj$ cross section is about a
factor~30 to~60 smaller than the $WWWjj$ cross section, depending on
the mass of the Higgs boson. We thus ignore the $WWZjj$ background in
the following.

\subsubsection{The $t\bar{t}j$ background}

We calculate $t\bar{t}j\to W^+bW^-\bar{b}j$ production where one of
the $b$-quarks decays semileptonically with an isolated lepton using
exact $t\bar{t}+$parton and $b\to c\ell\nu$ matrix elements. The
lepton from $b$-decay is considered not isolated if the charm quark is
within a cone of $\Delta R<0.4$ from the lepton and has
$p_T(c)>3$~GeV. Events are rejected with a factor two, which
approximates the fraction where the second $b$ quark would be tagged.
In order to regularize the soft parton $p_T$ distribution so as to
reproduce the $p_T$ distribution of the hard recoil system
($t\bar{t}$) from a full resummation calculation~\cite{tt}, while
preserving the normalization of the hard $t\bar{t}$ cross section, we
use the truncated shower approximation (TSA)~\cite{TSA}. The advantage
is that QCD matrix elements at tree level contain the full information
on angular distributions and hardness of additional jet emission. A
parton shower approach as used in Ref.~\cite{blondel} would not
immediately give reliable answers unless both color coherence and the
choice of scale are implemented correctly, matching the answer given
by QCD matrix elements for sufficiently hard partons. In practice this
is achieved by integrating the full tree-level $t\bar{t}+$parton
matrix elements over phase space down to $p_T({\rm parton}) > 1$~GeV,
and multiplying the result by a factor $1-\exp[-p_T^2({\rm
  parton})/p_{TSA}^2]$, where $p_{TSA}$ is adjusted to achieve
$\sigma_{t\bar{t}j} = \sigma_{t\bar{t}}$. For a scale choice of $\mu^2
= m_t^2$, we find $p_{TSA} = 15,\, 50$~GeV for the LHC and VLHC,
respectively.  This should not be construed as an attempt to mimic a
full NLO calculation of $t\bar{t}j$ production, which is not available, but it
does represent an improvement over using only {\sc pythia} $t\bar{t}$
matrix elements with additional partonic emission from showering, and
includes a well-motivated approach to controlling the soft singularity
present in the exact $\cal{O}$($\alpha_s^3)$ matrix elements.

\subsubsection{Other backgrounds: overlapping events and double parton
scattering}\label{sec:overlap}

At a high-luminosity intersection point of colliding beams, more than
one event may occur per bunch crossing. In principle, such overlapping
events can be recognized by a total visible energy measurement or by
tracing some final particle tracks back to distinct event vertices,
but in practice this may not always be possible. In this case, the
overlap of, e.g, two single $H\to W^+W^-$ events, a $W^+W^-$ and a
$Wjj$ event, or a $t\bar{t}$ and a single $W$ event may constitute a
potentially dangerous source of background for the Higgs boson pair
production signal. If the vertices of the overlapping events cannot be
resolved, the effective cross section for overlapping events is given
by~\cite{vernon}
\begin{equation}
\label{eq:over}
\sigma_{ov}(1,2)={1\over 2}\,\sigma(1)\,\sigma(2)\,{\cal L}_{bc},
\end{equation}
where $\sigma(1)$ and $\sigma(2)$ are the cross sections for the two
overlapping processes, and ${\cal L}_{bc}$ is the luminosity per bunch
crossing. It is given by
\begin{equation}
{\cal L}_{bc}={\cal L}\,\Delta\tau ,
\end{equation}
where ${\cal L}$ is the instantaneous luminosity and $\Delta\tau$ is
the bunch spacing. The values for ${\cal L}$, $\Delta\tau$ and ${\cal
  L}_{bc}$ at the LHC, SLHC and VLHC are listed in
Table~\ref{tab:one}.
\begin{table}
\caption{Luminosity, bunch spacing, and luminosity per bunch
crossing for the LHC~[\ref{SLHC1}], SLHC~[\ref{SLHC1}] and
VLHC~[\ref{VLHC1}].}  
\label{tab:one}
\vskip 3.mm
\begin{tabular}{cccc}
machine & luminosity ${\cal L}$ & bunch spacing $\Delta\tau$ & ${\cal
L}_{bc}$ \\
\tableline
LHC & $10^{34}\,{\rm cm^{-2}\,s^{-1}}$ & 25~ns & $(4.0~{\rm mb})^{-1}$ \\
SLHC & $10^{35}\,{\rm cm^{-2}\,s^{-1}}$ & 12.5~ns & $(0.8~{\rm mb})^{-1}$ \\
VLHC & $2\times10^{34}\,{\rm cm^{-2}\,s^{-1}}$ & 18.8~ns & $(2.7~{\rm
mb})^{-1}$
\end{tabular}
\end{table}

The $\ell^\pm{\ell'}^\pm jjjj$ final state can also be produced via
the independent scattering of two pairs of partons in the incident
protons. The cross section for double parton scattering is given by
Eq.~(\ref{eq:over}) with the factor ${\cal L}_{bc}/2$ replaced by
$1/\sigma_{\rm eff}$. The parameter $\sigma_{\rm eff}\approx
15$~mb~\cite{doublep}, the effective cross section, contains all the
information about the non-perturbative structure of the proton in this
simplified approach. It is believed that $\sigma_{\rm eff}$ is largely
independent of the center of mass energy~\cite{trele}. Comparison of
$\sigma_{\rm eff}$ and the values for ${\cal L}_{bc}$ listed in
Table~\ref{tab:one} shows that the double parton cross section is
about a factor~2 to~10 smaller than that from overlapping events.

\subsection{Numerical results}

The total cross sections within cuts (see Eqs.~(\ref{eq:cuts1})
--~(\ref{eq:cuts2})) for signal and background processes at the LHC
and VLHC are listed in Table~\ref{tab:two}.
\begin{table}
\caption{Higgs pair signal and background cross sections (fb) for
$pp\to\ell^\pm {\ell'}^\pm +4j$ ($\ell,\,\ell'=e,\,\mu$) at (a) the LHC
($\sqrt{s}=14$~TeV) and (b) at the VLHC ($\sqrt{s}=200$~TeV), imposing
the cuts listed in Eqs.~(\ref{eq:cuts1}) --~(\ref{eq:cuts2}), and as a 
function of the Higgs boson mass (GeV). 
The background labeled ``pileup'' represents a rough estimate of the
combined $WWWjj$, $t\bar{t}W$, $t\bar{t}Z$, $WZjjjj$, $WWjjjj$ and $t\bar
tt\bar{t}$ cross section from overlapping events
and double parton scattering. Cross sections at the SLHC are identical
to those in the LHC case with the exception of the pileup cross section,
which is about a factor~3.7 larger than at the LHC. The last column,
labeled ${\cal B}_{tot}$, shows the total background cross section.}
\label{tab:two}
\vskip 3.mm
\begin{tabular}{ccccccccccc}
\multicolumn{10}{c}{(a) LHC}\\
$m_H$ & $HH$ & $WWWjj$ & $t\bar{t}W$ & $t\bar{t}Z$ & $t\bar{t}j$ & $WZjjjj$ &
$WWjjjj$ & $t\bar{t}t\bar{t}$ & pileup & ${\cal B}_{tot}$ \\
\tableline
150 & 0.07 & 0.36 & 0.22 & 0.05 & 0.08 & 0.15 &
0.005 & 0.002 & $\sim 0.03$ & 0.90 \\
160 & 0.19 & 0.49 & 0.22 & 0.05 & 0.08 & 0.15 &
0.005 & 0.002 & $\sim 0.03$ & 1.03 \\
180 & 0.18 & 0.40 & 0.22 & 0.05 & 0.08 & 0.15 &
0.005 & 0.002 & $\sim 0.03$ & 0.94 \\
200 & 0.08 & 0.29 & 0.22 & 0.05 & 0.08 & 0.15 &
0.005 & 0.002 & $\sim 0.03$ & 0.83 \\
\tableline
\multicolumn{10}{c}{(b) VLHC}\\
$m_H$ & $HH$ & $WWWjj$ & $t\bar{t}W$ & $t\bar{t}Z$ & $t\bar{t}j$ & $WZjjjj$ &
$WWjjjj$ & $t\bar{t}t\bar{t}$ & pileup & ${\cal B}_{tot}$ \\
\tableline
140 & 2.2 & 14.9 & 5.8 & 7.4 & 7.7 & 8.1 &
0.13 & 6.13 & $\sim 20$ & 70.2 \\
150 & 6.5 & 17.0 & 5.8 & 7.4 & 7.7 & 8.1 &
0.13 & 6.13 & $\sim 20$ & 72.3 \\
160 & 15.8 & 20.4 & 5.8 & 7.4 & 7.7 & 8.1 &
0.13 & 6.13 & $\sim 20$ & 75.7 \\
180 & 16.0 & 17.9 & 5.8 & 7.4 & 7.7 & 8.1 &
0.13 & 6.13 & $\sim 20$ & 73.2 \\
200 & 8.7 & 14.3 & 5.8 & 7.4 & 7.7 & 8.1 &
0.13 & 6.13 & $\sim 20$ & 69.6
\end{tabular}
\end{table}
At the LHC, with 300~fb$^{-1}$, at most about 50~signal events are
produced. Outside of the Higgs boson mass range considered here, the
number of signal events is too small to be useful. For $m_H<150$~GeV,
this is due to the small $H\to W^*W$ branching ratio. For
$m_H>200$~GeV, the $gg\to HH$ cross section is too small. $WWWjj$ and
$t\bar{t}W$ production are the largest contributions to the
background. The background from $t\bar{t}Z$ production where one of
the leptons is lost is moderate. Although the cross section for
$WZjjjj$ production is substantial, this background can be separated
rather easily from the signal as discussed below.  $WWjjjj$ and
$t\bar{t}t\bar{t}$ production contribute negligibly to the background
at the LHC. The $t\bar{t}t\bar{t}$ cross section is suppressed by the
large top quark mass. The $WWjjjj$ cross section is small because
quark-gluon and gluon-gluon fusion processes do not contribute to same
sign $W$ pair production.

The $t\bar{t}j$ cross section is extremely sensitive to the lepton
$p_T$ cut imposed. Requiring $p_T(\ell)>15$~GeV, the $t\bar{t}j$
background is of the same size as the $t\bar{t}Z$ background, and
about a factor~3 smaller than the $t\bar{t}W$ background. Decreasing
the $p_T(\ell)$ cut to 10~GeV, the $t\bar{t}j$ cross section increases
by about a factor~10, overwhelming the Higgs pair signal\footnote{In
  Ref.~\cite{BPR} we required one lepton with a $p_T>10$~GeV, and one
  lepton with $p_T>15$~GeV. In this case, the $t\bar{t}j$ background is
  significantly larger than the signal (see also Ref.~\cite{blondel}).
  Increasing the lepton transverse momentum cut to 15~GeV solves this
  problem, while reducing the signal cross section by less than
  $10\%$.}. On the other hand, if the minimum lepton transverse
momentum is increased to 20~GeV which reduces the signal cross section
by about $20\%$, the $t\bar{t}j$ background decreases by one order of
magnitude and essentially becomes negligible. However, we emphasize
that our matrix element based calculation of the $t\bar{t}j$
background should be viewed with some caution. Effects from
hadronization, event pileup and extra jets from initial or final state
radiation, as well as detector resolution effects may significantly
affect the cross section. For a reliable estimate of the background, a
full detector simulation, which is beyond the scope of this paper, is
required.

The lepton isolation requirement, together with the lepton $p_T$ cut,
the $b\to c\ell\nu$ branching ratio and the di-jet invariant mass cut
suppress the $t\bar{t}j$ cross section by about a factor $10^6$. A
similar suppression factor is also expected in $\ell^\pm\nu
b\bar{b}+3j$ production which also contributes to the background if
one of the $b$-quarks decays semileptonically and if the lepton from
$b$-decay is isolated. Using the result of Ref.~\cite{alpgen}, we
estimate that the $\ell^\pm\nu b\bar{b}+3j$ cross section at the LHC
is of ${\cal O}(10^{-3}$~fb) which can safely be neglected.

Our numerical results for the overall normalization of the signal, the
$WWWjj$, $t\bar{t}W$, and the $t\bar{t}t\bar{t}$ background processes
agree reasonably well with those reported in Refs.~\cite{SLHC}
and~\cite{blondel}. For $WZjjjj$ production, we find a cross section
which is about a factor~10 larger. The discrepancy can be traced to
the contribution from virtual photon exchange, which was not taken
into account in~\cite{SLHC,blondel}. No result for $t\bar{t}Z$
production is given in Refs.~\cite{SLHC} and~\cite{blondel}. A
meaningful comparison of our matrix element based calculation of the
$t\bar{t}j$ background and the {\sc pythia} based estimate
in~\cite{SLHC,blondel} is not possible due to the strong dependence of
the cross section on the lepton $p_T$.

Overlapping events and double parton scattering are not expected to
contribute significantly to the background at the LHC. Contributions
from these sources are listed in Table~\ref{tab:two} in the column
labeled as ``pileup''. The numerical values listed were obtained by
adding all overlapping event and double parton scattering
contributions to the sources of background discussed in
Sec.~\ref{sec:bgd}, using Eq.~(\ref{eq:over}) and the values of ${\cal
  L}_{bc}$ given in Table~\ref{tab:one}. Since Eq.~(\ref{eq:over})
assumes that that the vertices of the two overlapping events are not
resolved, these values are likely conservative. For the SLHC, the
pileup cross section in Table~\ref{tab:two}a has to be multiplied by a
factor~3.7. Our results for the cross section from overlapping events
and double parton scattering should be regarded only as order of
magnitude estimates. Realistic simulations are needed to draw firm
conclusions for this background.

At a $pp$ collider with $\sqrt{s}=200$~TeV, the cross sections of
processes which are dominated by gluon fusion, such as the the $gg\to
HH$ signal, $t\bar{t}t\bar{t}$, $t\bar{t}Z$ and $t\bar{t}j$ production,
are about a factor~100 --~3000 larger than that at the LHC. Since the
cross section for $t\bar{t}t\bar{t}$ production at the LHC is suppressed
by the large invariant mass of four top quarks, the increase is
particularly large for this process. In contrast, the cross sections
of processes dominated by quark-gluon fusion or quark-quark
scattering, such as $WWWjj$, $t\bar{t}W$ and $WWjjjj$ production,
increase by only a factor~25 --~45. As a result, the $t\bar{t}Z$,
$t\bar{t}j$ and $t\bar{t}t\bar{t}$ backgrounds are relatively more
important at the VLHC. The cross sections due to overlapping events
and double parton scattering increase by almost three orders of
magnitude, and thus may well compete in size with $WWWjj$ production,
unless the vertex positions of the overlapping events are resolved.
Since the signal is purely gluon induced, the overall signal to
background ratio at the VLHC is about a factor~2 better than at the
LHC.

\begin{figure}[h!] 
\begin{center}
\includegraphics[width=13.3cm]{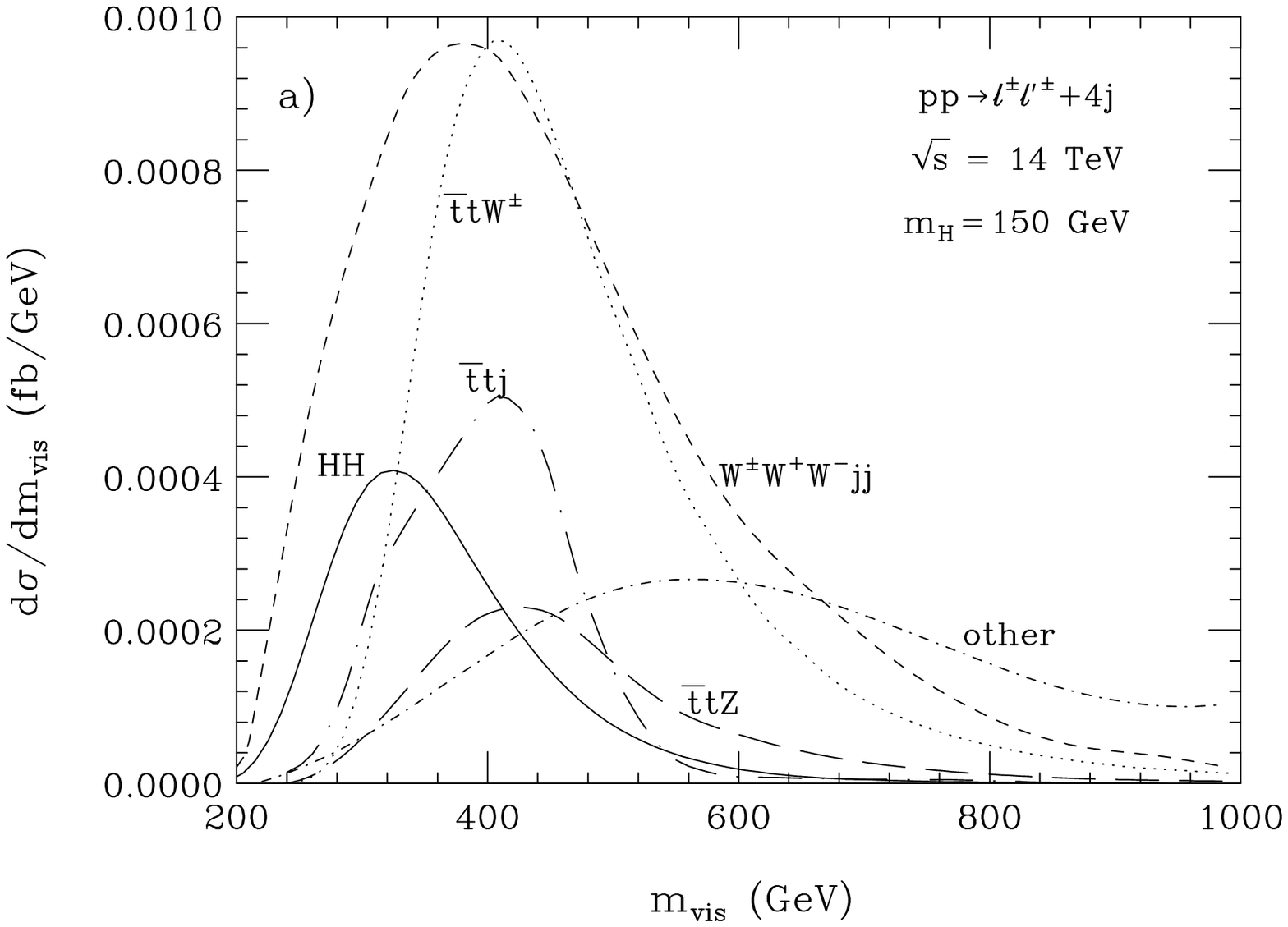} \\[3mm]
\includegraphics[width=13.3cm]{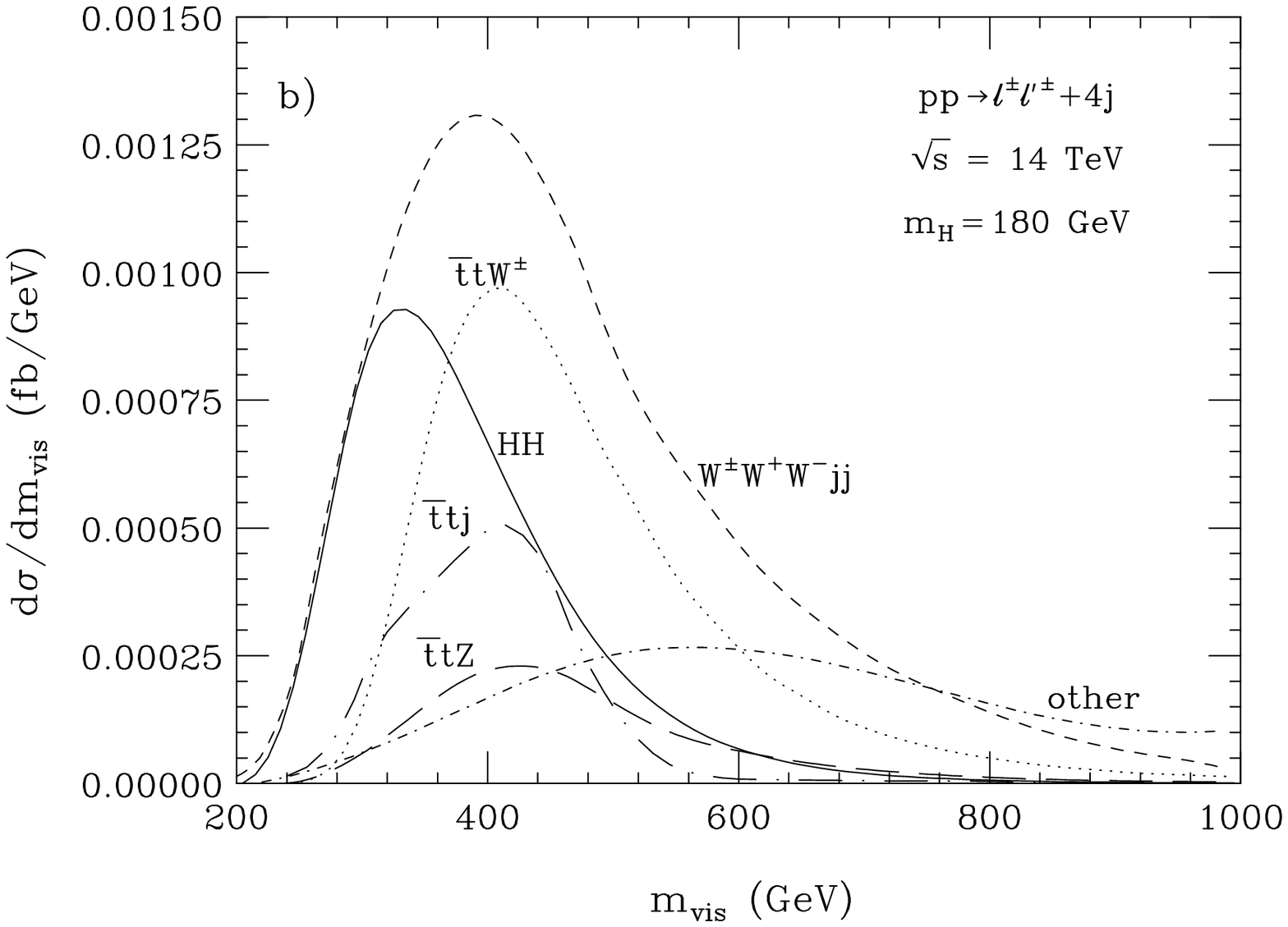}
\vspace*{2mm}
\caption[]{\label{fig:fig4} 
  Distribution of the invariant mass of the observable final state
  particles, $m_{vis}$, after all cuts, in
  $pp\to\ell^\pm{\ell'}^\pm+4j$ for the signal with a) $m_H=150$~GeV
  and b) $m_H=180$~GeV, and all backgrounds (except for the
  contributions from overlapping events and double parton scattering)
  at the LHC. The dot-dashed curve shows the combined cross section of
  $WZjjjj$, $WWjjjj$ and $t\bar{t}t\bar{t}$ production.}
\vspace{-7mm}
\end{center}
\end{figure}

\begin{figure}[h!] 
\begin{center}
\includegraphics[width=13.cm]{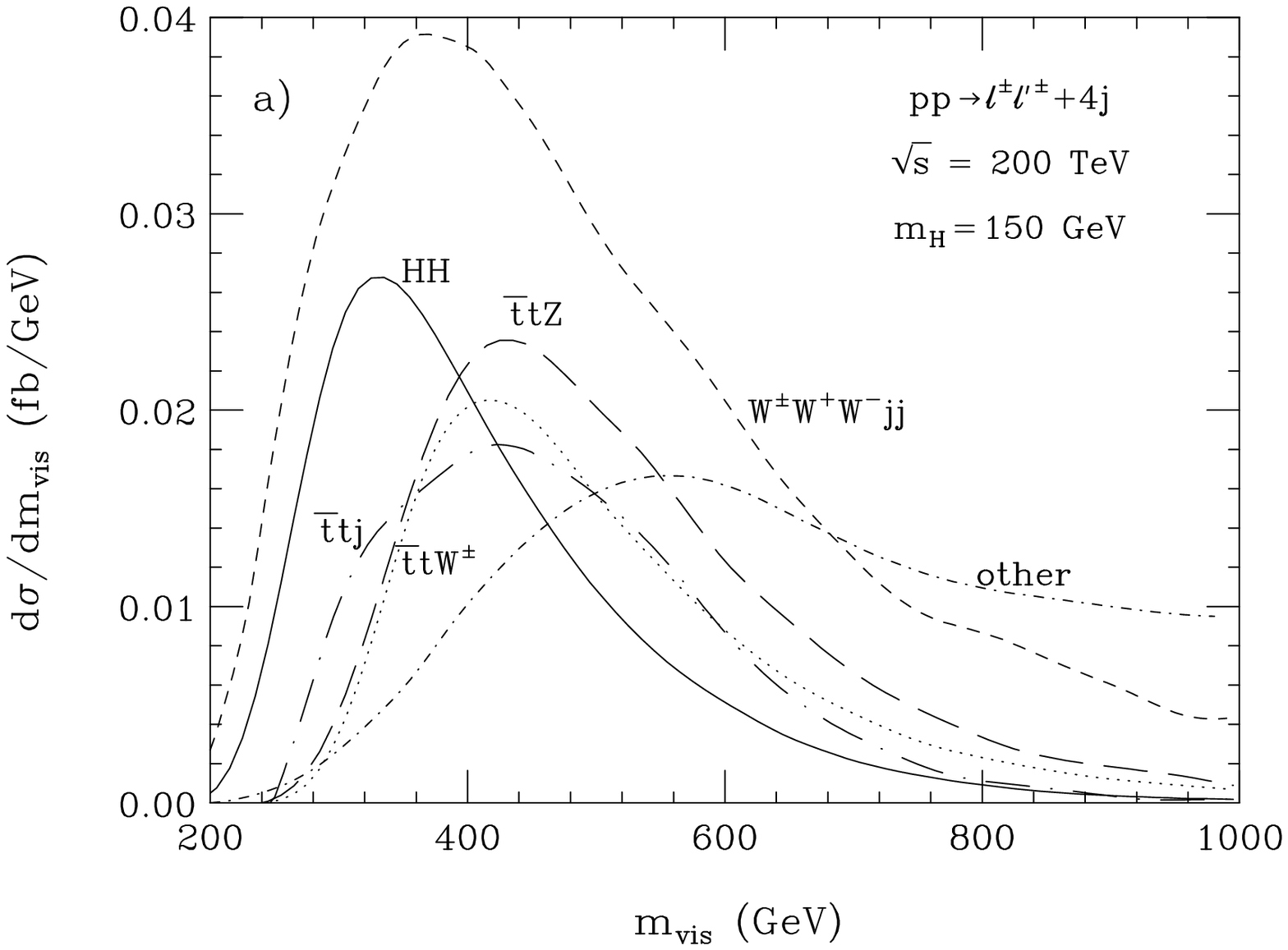} \\[3mm]
\includegraphics[width=13.cm]{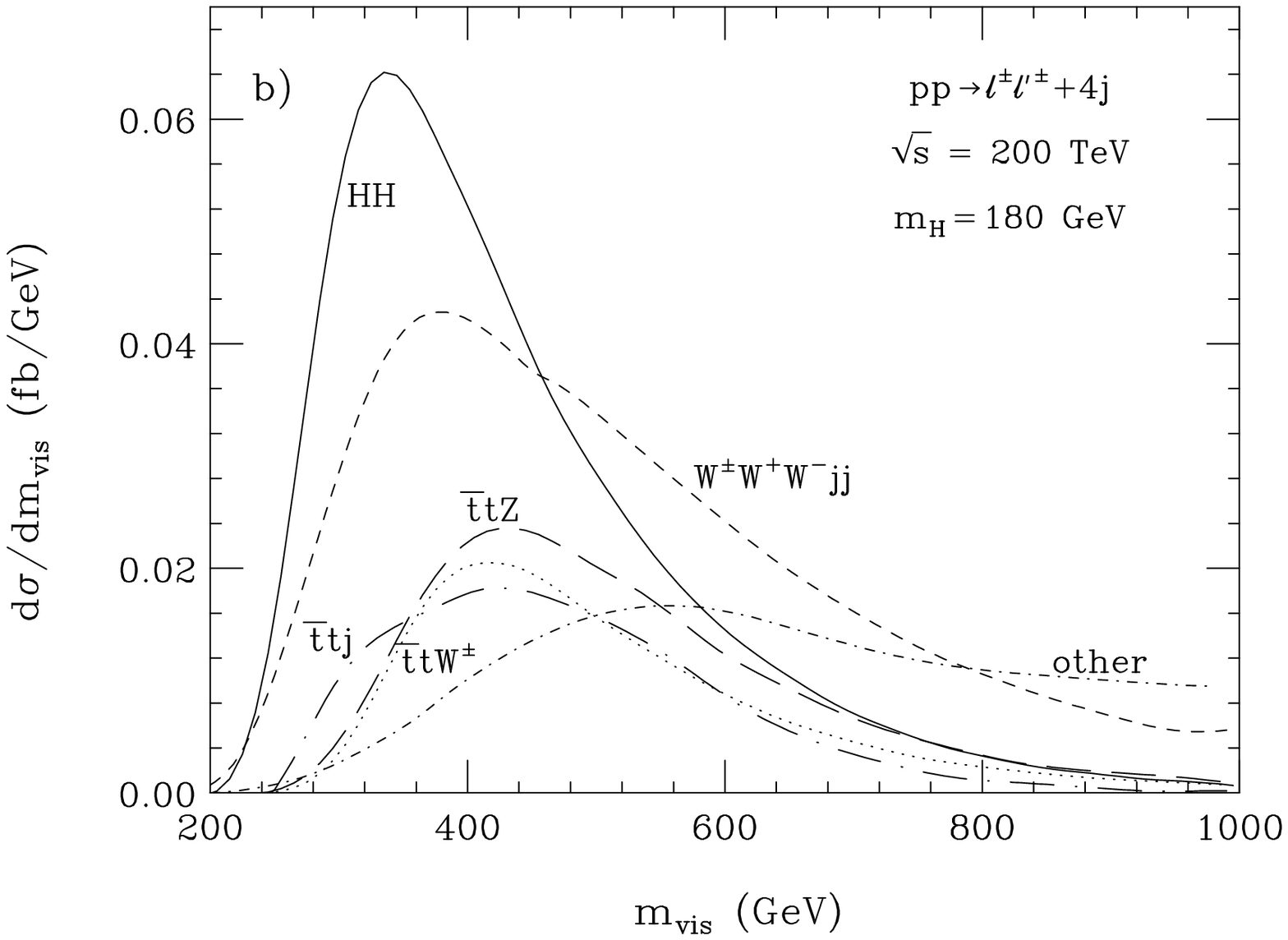}
\vspace*{2mm}
\caption[]{\label{fig:fig5} 
  Distribution of the invariant mass of the observable final state
  particles, $m_{vis}$, after all cuts, in
  $pp\to\ell^\pm{\ell'}^\pm+4j$ for the signal (solid line) with a)
  $m_H=150$~GeV and b) $m_H=180$~GeV, and all backgrounds (except for
  the contributions from overlapping events and double parton
  scattering) at the VLHC (dashed: $WWWjj$, dotted: $t\bar{t}W$,
  long-dashed: $t\bar{t}Z$, long-dash-dot: $t\bar{t}j$).  The
  dot-dashed curve shows the combined cross section of $WZjjjj$,
  $WWjjjj$ and $t\bar{t}t\bar{t}$ production.}  \vspace{-7mm}
\end{center}
\end{figure}

All the backgrounds are multi-body production processes, therefore the
distribution of the invariant mass, $\sqrt{\hat s}$, of the system
peaks at values significantly above threshold. In contrast, the signal
is a two-body production process for which the $\sqrt{\hat s}$
distribution will exhibit a sharper threshold behavior. Unfortunately,
with two neutrinos present in the final state, $\sqrt{\hat s}$ cannot
be reconstructed. However, we anticipate that the invariant mass of
all observed final state leptons and jets given by ($E_i$ and
${\mathbf p}_i$ are the energies and momenta of the jets and leptons)
\begin{equation}
\label{eq:mvis}
m^2_{vis}=\left[\sum_{i=\ell,{\ell'},\,{\rm jets}}E_i\right]^2 -
\left[\sum_{i=\ell,{\ell'},\,{\rm jets}}\mathbf p_i\right]^2
\end{equation}
will retain most of the expected behavior of the different production
processes. Figures~\ref{fig:fig4} (LHC) and~\ref{fig:fig5} (VLHC)
clearly demonstrate that this is the case: the signal peaks at smaller
values of $m_{vis}$ than the background processes, especially for
lower Higgs boson masses. 
This distribution, which was not considered
in Ref.~\cite{SLHC}, is what makes possible a $\chi^2$ based test to
improve extraction of the Higgs boson self-coupling (see
Sec.~\ref{sec:sec4}). While detector effects may smear out the tails
of this distribution, or shift a peak slightly, it is a genuine,
simple physics effect and care must be taken in any approximations
used to simulate the backgrounds that this physics feature is
retained.

Since the $WWWjj$ background has a significant contribution from
$WH(\to W^+W^-)jj$ production, its $m_{vis}$ distribution is similar
in shape to that of the $HH$ signal. As expected, the $m_{vis}$
distributions of $t\bar{t}W$, $t\bar{t}Z$ and $t\bar{t}j$ production
peak at similar values, and are similar in shape. The dot-dashed lines
in Figs.~\ref{fig:fig4} and~\ref{fig:fig5} represent the combined
differential cross section of $WZjjjj$, $WWjjjj$ and $t\bar{t}t\bar{t}$
production. It peaks at a much higher visible invariant mass than
those of the other background processes. Whereas the signal is
concentrated in the region $m_{vis}<500$~GeV, the background processes
have a significant tail extending to $m_{vis}=1$~TeV and beyond. This
makes it possible to normalize the background using data from the
$m_{vis}>500$~GeV region. The simple procedure described in
Sec.~\ref{sec:overlap} for estimating the cross section for
overlapping events and double parton scattering is not suitable for
calculating distributions. The $m_{vis}$ distribution of these
backgrounds therefore is not included in Figs.~\ref{fig:fig4}
and~\ref{fig:fig5}. With increasing Higgs boson mass, the signal peak
gradually moves to higher values of $m_{vis}$. The efficiency of the
$m_{vis}$ distribution as a discriminator thus decreases somewhat for
$m_H>180$~GeV. Comparing the visible invariant mass distributions at
the LHC and VLHC, the improved signal to background ratio at the VLHC
in the region $m_{vis}=200-500$~GeV, where the Higgs pair signal is
concentrated, is obvious.

As noted before, all our calculations are consistently performed at
leading order, ie. there are precisely four jets (partons) in the
final state. In practice, one expects a significant fraction of the
$\ell^\pm{\ell'}^\pm+4j$ signal events to contain one or more extra
jets originating from initial state gluon radiation. In such events,
it is natural to construct $m_{vis}$ from the four highest $p_T$ jets
in the event. However, there is no guarantee that the extra jets are
always the softest jets in the event. Since the $m(jj)$ requirement is
rather loose (see Eq.~(\ref{eq:jjmass})), it is conceivable that
events where one (or more) of the four jets incorporated in $m_{vis}$
originate from QCD bremsstrahlung. Hard QCD corrections could also
lead to $\ell^\pm{\ell'}^\pm +4j$ events where one of the jets from
$W$ decay does not pass the minimum $p_T$ cut for jets, but the
additional bremsstrahlung jet does. Some of these events might also
pass the $m(jj)$ cut. QCD corrections thus could affect the shape of
the $m_{vis}$ distribution. In order to draw firm conclusions, a full
calculation of the NLO QCD corrections to $gg\to HH$ with finite top
quark mass effects is needed. Insight may also be gained from
performing a calculation where the $gg\to HH$ matrix elements are
interfaced~\cite{lafaye} with an event generator such as {\sc pythia}.

In using {\sc pythia} for the additional jet radiation, however, one
has to be careful. As described previously, the radiation of soft and
collinear jets from the initial state is the main source of the large
(and top mass independent) QCD corrections to the total signal cross
section. The initial state radiation modeled by {\sc pythia}
effectively resums the leading effects of precisely this radiation and
includes it in the topology of the final state. Normalizing the rate
to the leading order total cross section is therefore inconsistent
and the result arbitrary (and not, as often is claimed, a conservative
estimate), because the final state topology and the rate are computed
in different approximations with a difference which is by no means a
reduced higher order uncertainty.

The effect of hard QCD corrections on the $m_{vis}$ distribution may
be reduced by limiting the number of possible $4j$ combinations which
satisfy the cut of Eq.~(\ref{eq:jjmass}). Approximately $60-65\%$
($35-40\%$) of all signal events have one (two) $4j$ combination
satisfying Eq.~(\ref{eq:jjmass}); almost none have three $4j$
combinations in the correct invariant mass range. When additional jets
are present, many more combinations are possible. Adding the
requirement that at most two $4j$ combinations satisfy
Eq.~(\ref{eq:jjmass}) may thus reduce the effect of hard QCD
corrections on the $m_{vis}$ distribution. It may also reduce the
signal and background cross sections somewhat.  The fraction of events
where one or several QCD bremsstrahlung jets pass the cuts may also be
reduced by shrinking the $m(jj)$ range in Eq.~(\ref{eq:jjmass}) (see
also Ref.~\cite{blondel}). Our choice has been deliberately
conservative. Reducing the di-jet invariant mass range to $M_W\pm
20$~GeV may well be possible~\cite{lhctop}. This would also improve
the signal to background ratio.

\section{The three lepton final state}
\label{sec:sec3}

The calculation of signal and background cross sections for the
$(jj\ell^\pm\nu)({\ell'}^\pm\nu {\ell''}^\mp\nu)$ final state is
similar to that described in Sec.~\ref{sec:sec2} for the same sign
di-lepton final state. Due to the smaller branching ratio for leptonic
$W$ decays, the cross section is expected to be somewhat smaller than
that for the $(jj\ell^\pm\nu)(jj{\ell'}^\pm\nu)$ channel.  The
kinematic acceptance cuts for both signal and backgrounds in the
$(jj\ell^\pm\nu)({\ell'}^\pm\nu {\ell''}^\mp\nu)$ final state are:
\begin{eqnarray}
\label{eq:cuts3}
&p_T(j) > 30,\,20~{\rm GeV} , \qquad 
p_T(\ell) > 15,\,15,\,15~{\rm GeV}    ,        \\
&|\eta(j)| < 3.0     ,        \qquad \qquad \qquad \qquad 
|\eta(\ell)| < 2.5   ,              \\
&\Delta R(jj) > 0.6 ,   \qquad 
\Delta R(j \ell) > 0.4 , \qquad 
\Delta R(\ell \ell) > 0.2 .
\label{eq:cuts4}
\end{eqnarray}
In addition, we impose the di-jet invariant mass cut of
Eq.~(\ref{eq:jjmass}). 

Except for $W^\pm W^\pm jjjj$ and $W^\pm Zjjjj$ production, all
processes discussed in Sec.~\ref{sec:bgd} contribute to the
background. For final states containing a same flavor opposite sign
lepton pair, $W^\pm \ell^+\ell^- jj$ production constitutes an
additional source of background.  We have calculated the $W^\pm
\ell^+\ell^- jj$ background using the exact matrix elements of
Ref.~\cite{vernon1}. The cross section for $W^\pm \ell^+\ell^- jj$
production is about three orders of magnitude larger than the Higgs
boson pair signal, if photon exchange diagrams are taken into account.
To increase the signal to background ratio, one can either impose a
minimum $\ell^+\ell^-$ invariant mass cut, or increase the $\Delta
R(\ell^+ \ell^-)$ cut. Unfortunately, due to correlations between the
momenta of the fermions in the $H\to W^+W^-\to 4$~fermion
decay~\cite{glover}, the $\ell^+\ell^-$ invariant mass tends to be
rather small in Higgs pair events. As a result, the signal to
background ratio cannot be improved to better than 1:100 without
reducing the signal cross section to an unacceptably low level.

In the following we therefore only consider
$\ell^\pm{\ell'}^\mp{\ell'}^\mp+2j$, $\ell\neq\ell'$, production.
Excluding all final states containing a same flavor opposite sign
lepton pair reduces the signal cross section by a factor~4. As a
result, the cross section at the LHC becomes too small to be of
interest; for an integrated luminosity of 300~fb$^{-1}$ only 8~events
are expected. We therefore present numerical results at VLHC energies
only in this section.

\begin{figure}[h!] 
\begin{center}
\includegraphics[width=13.cm]{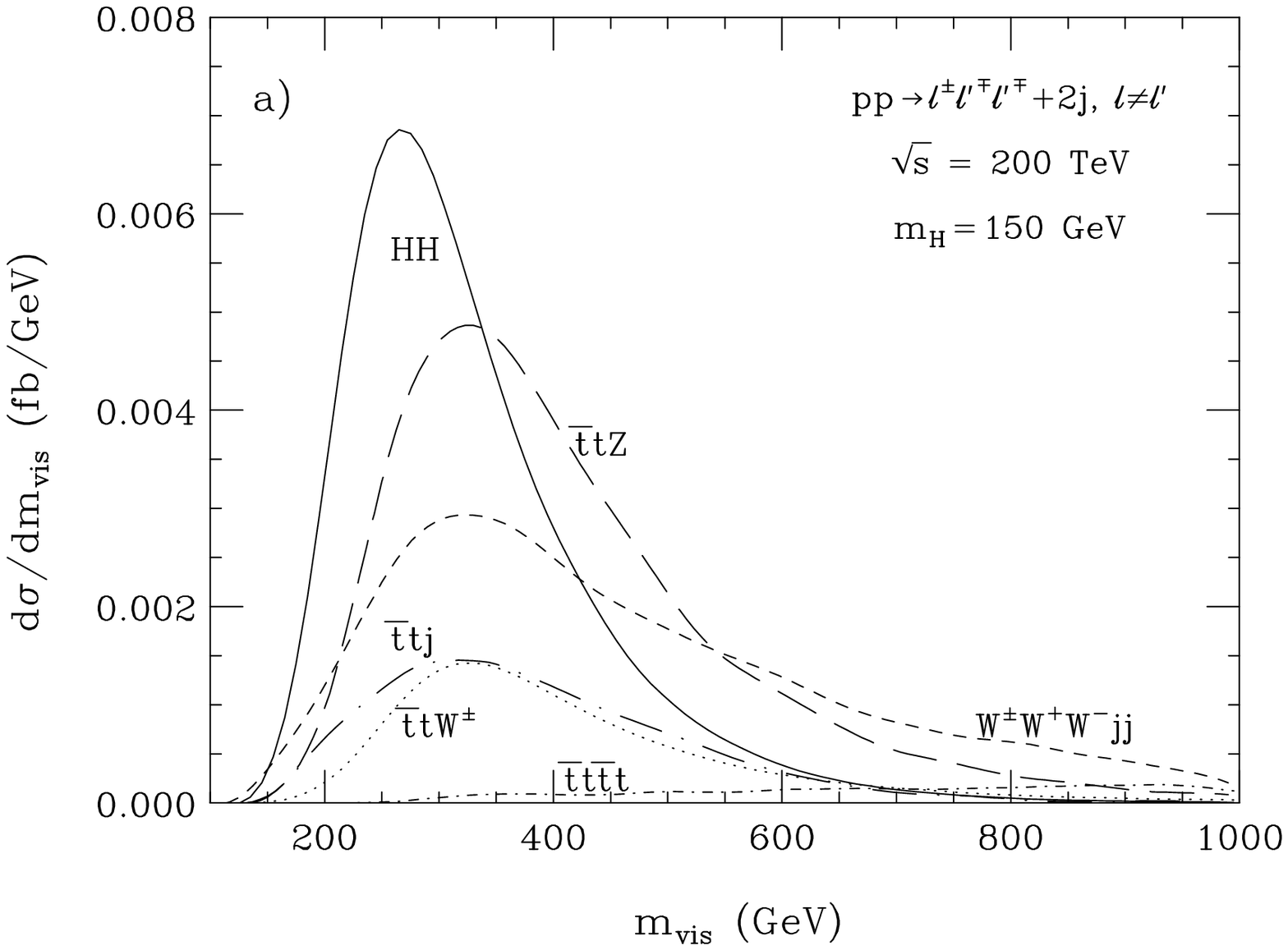} \\[3mm]
\includegraphics[width=13.cm]{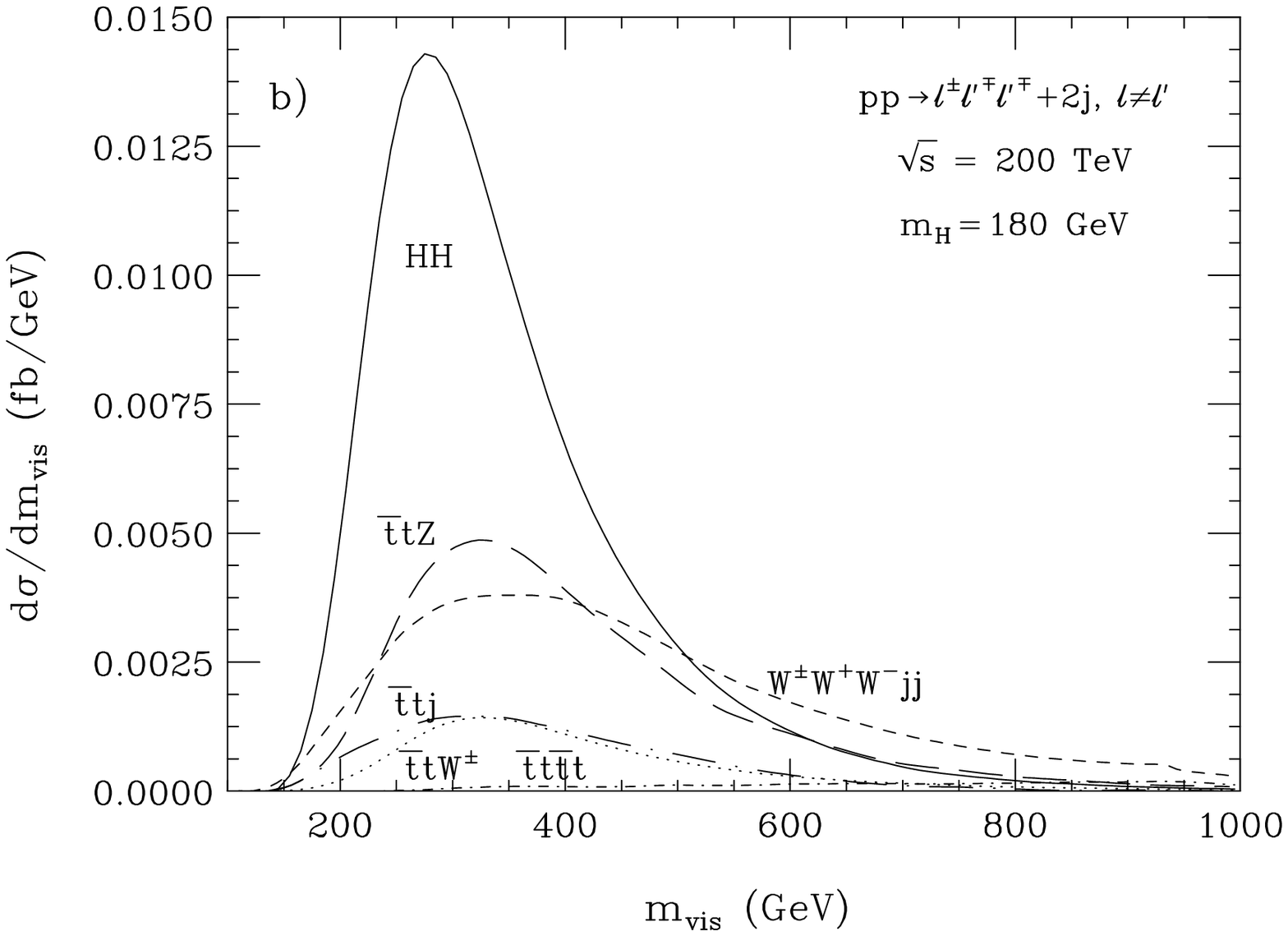}
\vspace*{2mm}
\caption[]{\label{fig:fig6} 
  Distribution of the invariant mass of the observable final state
  particles, $m_{vis}$, after all cuts, in
  $pp\to\ell^\pm{\ell'}^\mp{\ell'}^\mp+2j$, $\ell\neq\ell'$, for the
  signal (solid line) with a) $m_H=150$~GeV and b) $m_H=180$~GeV, and
  all backgrounds (except for the contributions from overlapping
  events and double parton scattering) at the VLHC (dashed: $WWWjj$,
  dotted: $t\bar{t}W$, long-dashed: $t\bar{t}Z$, long-dash-dot:
  $t\bar{t}j$, dot-dashed: $t\bar{t}t\bar{t}$ ). }  \vspace{-7mm}
\end{center}
\end{figure}

The total cross sections within cuts (see Eqs.~(\ref{eq:cuts3})
--~(\ref{eq:cuts4})) for signal and background processes at the VLHC
are listed in Table~\ref{tab:three}.
\begin{table}
\caption{Higgs pair signal and background cross sections (fb) for
$pp\to\ell^\pm{\ell'}^\mp{\ell'}^\mp+2j$
($\ell,\,\ell'=e,\,\mu$, $\ell\neq\ell'$) at the VLHC
($\sqrt{s}=200$~TeV), imposing 
the cuts listed in Eqs.~(\ref{eq:cuts3}) --~(\ref{eq:cuts4}), and
as a function of the Higgs boson mass (GeV).
The background labeled ``pileup'' represents a rough estimate of the
combined $WWWjj$, $t\bar{t}W$, $t\bar{t}Z$, and $t\bar
tt\bar{t}$ cross section from overlapping events
and double parton scattering. The last column,
labeled ${\cal B}_{tot}$, shows the total background cross section.}
\label{tab:three}
\vskip 3.mm
\begin{tabular}{ccccccccc}
$m_H$ & $HH$ & $WWWjj$ & $t\bar{t}W$ & $t\bar{t}Z$ & $t\bar{t}j$ & 
$t\bar{t}t\bar{t}$ & pileup & ${\cal B}_{tot}$ \\
\tableline
150 & 1.40 & 1.43 & 0.39 & 1.39 & 0.45 & 0.47 &
$\sim 2.4$ & 6.53 \\
160 & 3.06 & 1.96 & 0.39 & 1.39 & 0.45 & 0.47 &
$\sim 2.4$ & 7.06 \\
180 & 3.04 & 1.71 & 0.39 & 1.39 & 0.45 & 0.47 &
$\sim 2.4$ & 6.81 \\
200 & 1.66 & 1.47 & 0.39 & 1.39 & 0.45 & 0.47 &
$\sim 2.4$ & 6.57
\end{tabular}
\end{table}
The signal cross section is about a factor~5 smaller than in the
same-sign dilepton case. The loss in signal however is at least
partially compensated by the significantly improved signal to
background ratio. This becomes more evident in Fig.~\ref{fig:fig6}
which shows the $m_{vis}$ distribution for $m_H=150$~GeV and
$m_H=180$~GeV.
Here, $m_{vis}$ is defined by Eq.~(\ref{eq:mvis}), using the
four-momentum vectors of the three charged leptons and the two jets.
The largest contribution to the background in the region where the
signal peaks comes from $t\bar{t}Z$ production. While the $t\bar{t}j$
and $t\bar{t}t\bar{t}$ total cross sections are similar (see
Table~\ref{tab:three}), the $t\bar{t}t\bar{t}$ contribution to the
background in this region is negligible. Most $t\bar{t}t\bar{t}$
events have visible invariant masses well in excess of 1~TeV. The
signal to background ratio for the $\ell^\pm{\ell'}^\mp{\ell'}^\mp+2j$
channel is approximately a factor two better than for the same-sign
dilepton final state (see Fig.~\ref{fig:fig5}).

Since there are fewer jets present, initial state gluon radiation
should have a smaller effect on the $m_{vis}$ distribution in the
three lepton final state than in the same-sign dilepton case.

\section{Determining the Higgs Boson Self-Coupling}
\label{sec:sec4}

\begin{figure}[h!] 
\begin{center}
\includegraphics[width=13.cm]{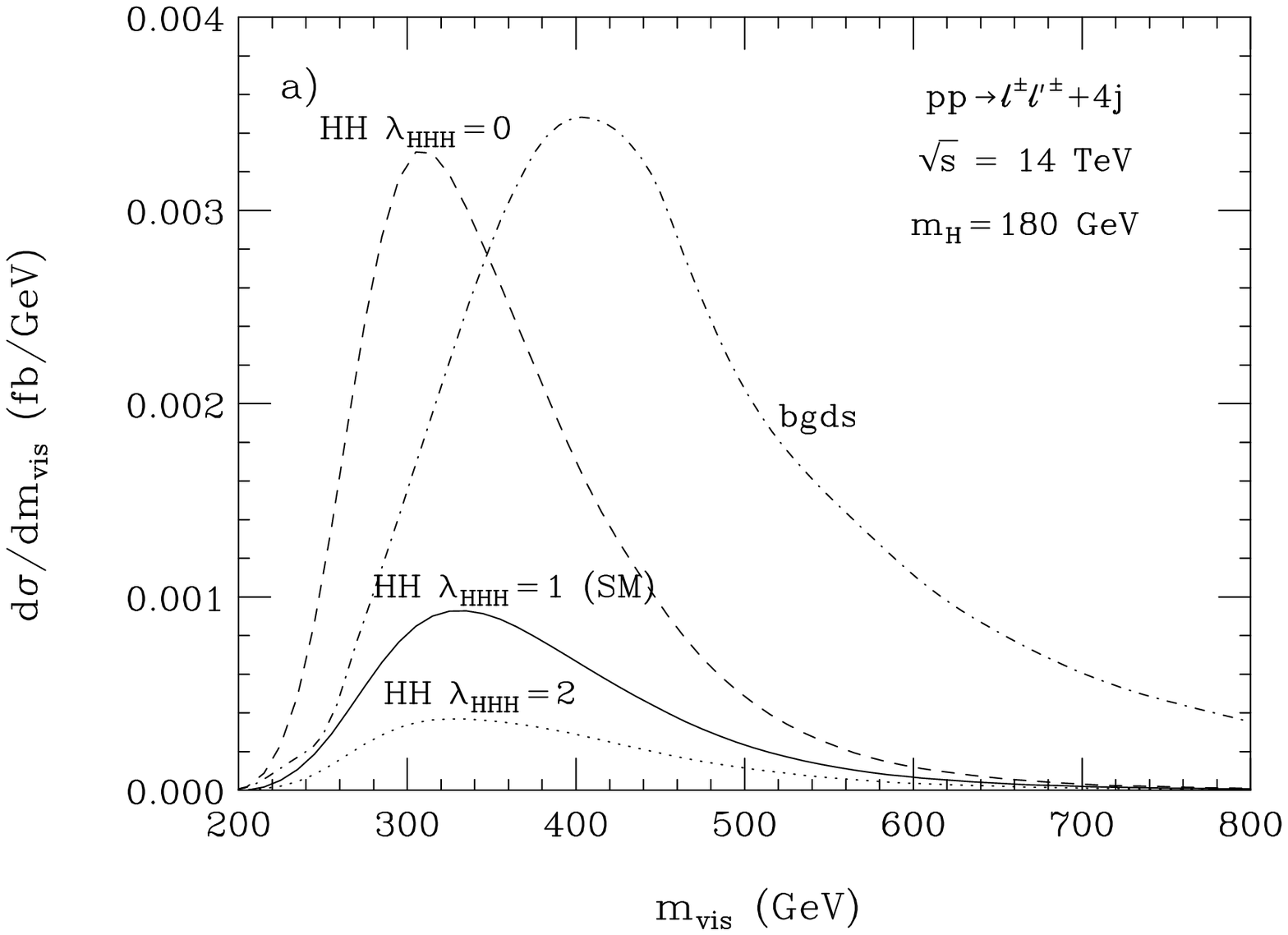} \\[3mm]
\includegraphics[width=13.cm]{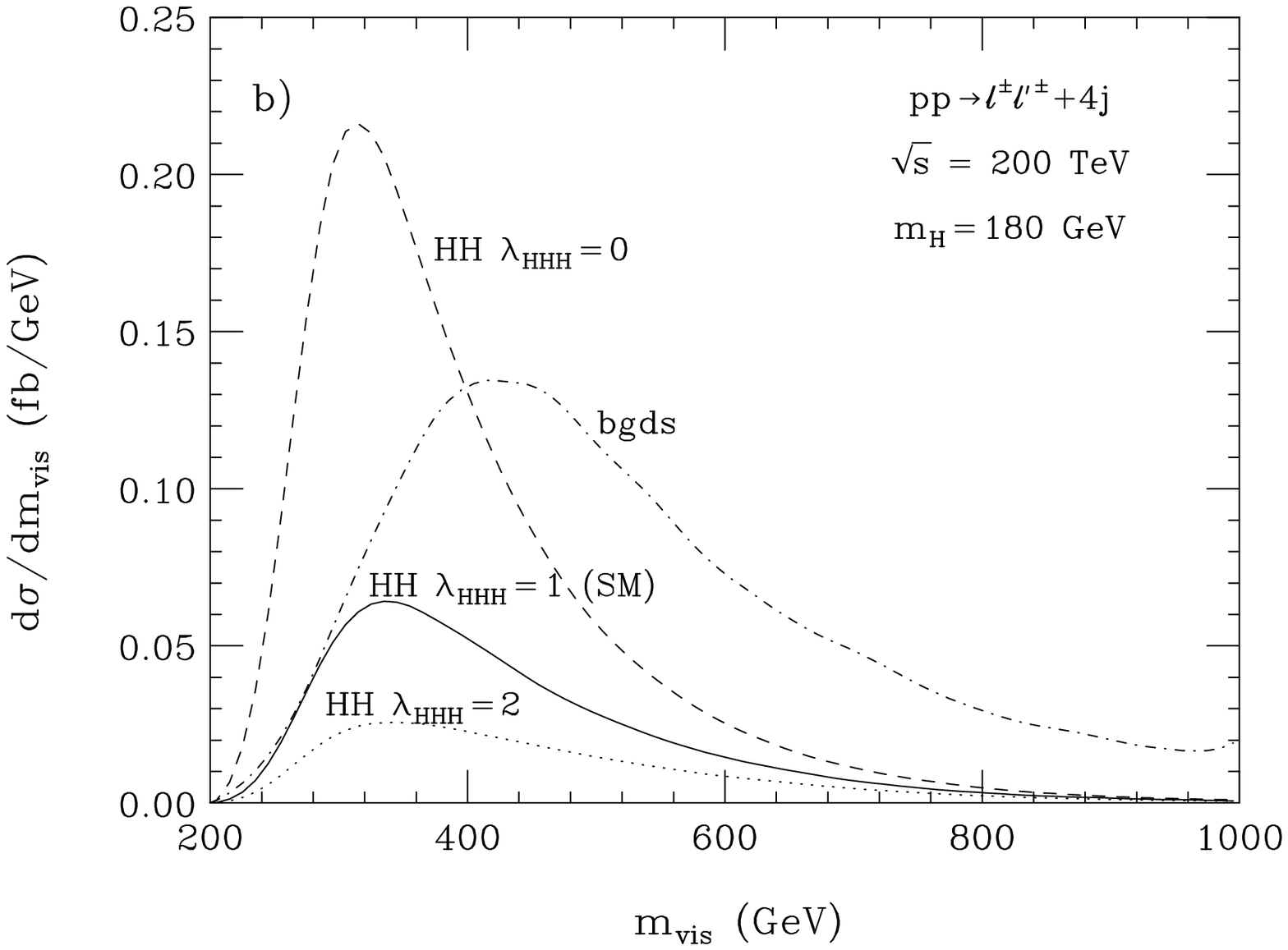}
\vspace*{2mm}
\caption[]{\label{fig:fig7} 
  The $m_{vis}$ distribution of the signal for $pp\to
  \ell^\pm{\ell'}^\pm+4j$ and $m_H=180$~GeV at a) the LHC, and b) the
  VLHC, in the SM (solid curve), for
  $\lambda_{HHH}=\lambda/\lambda_{SM}=0$ (dashed line) and for
  $\lambda_{HHH}=2$ (dotted line).  The dot-dashed line shows the
  combined $m_{vis}$ distribution of all background processes (except
  those from overlapping events and double parton scattering).
  Qualitatively similar results are obtained for other values of
  $m_H$. }  \vspace{-7mm}
\end{center}
\end{figure}

The Feynman diagrams contributing to $gg\to HH$ in the SM consist of
fermion triangle and box diagrams~\cite{higgs_self}.  Non-standard
Higgs boson self-couplings affect only the triangle diagrams with a
Higgs boson exchanged in the $s$-channel. They contribute only to the
$J=0$ partial wave, and thus impact the $m_{vis}$ distribution mostly
at small values. This is illustrated in Fig.~\ref{fig:fig7} for the
$\ell^\pm{\ell'}^\pm+4j$ final state with $m_H=180$~GeV and two
non-standard values of $\lambda_{HHH}=\lambda/\lambda_{SM}$.
Since box and triangle diagrams interfere destructively, the $gg\to
HH$ cross section for $1<\lambda_{HHH}<2.7$ is smaller than in the SM.
The absence of a Higgs boson self-coupling ($\lambda_{HHH}=0$) results
in a Higgs pair production cross section which is about a factor~3
larger than the SM result. Figure~\ref{fig:fig7} also demonstrates
that the $m_{vis}$ distribution of the signal peaks at a smaller value
than that of the combined background. This remains true for other
Higgs boson masses, as long as $m_H\leq 200$~GeV.

The shape change of the $m_{vis}$ distribution induced by non-standard
values of $\lambda_{HHH}$ can be used to derive quantitative
sensitivity bounds on the Higgs boson self-coupling. We accomplish
this by calculating $95\%$ confidence level (CL) limits performing a
$\chi^2$ test. The statistical significance is calculated by splitting
the $m_{vis}$ distribution into a number of bins, each with typically
more than five events. In each bin the Poisson statistics are
approximated by a Gaussian distribution. We impose the cuts described
in Secs.~\ref{sec:sec2} and~\ref{sec:sec3} and combine channels with
electrons and muons in the final state, conservatively assuming a
common lepton identification efficiency of $\epsilon=0.85$ for each
lepton. Except for the Higgs boson self-coupling we assume the SM to
valid: by the time a measurement of $\lambda$ is attempted, the Higgs
boson mass will be precisely known and the $H\to W^+W^-$ branching
ratio will have been measured with a precision of $10\%$ or better at
the LHC and/or an $e^+e^-$ linear collider~\cite{zepp}. We include all
background processes listed in Table~\ref{tab:two} and
Table~\ref{tab:three}, except those from overlapping events and double
parton scattering. The challenge of including higher order effects is
considerably more complicated for the background than for the $HH$
signal, where at least the physics interpretation is clear as
previously discussed. The aim for the backgrounds is not to capture
the bulk of events after cuts. Instead, one tries to cut into the
tails of distributions, where the impact of higher order corrections
might be very different. Therefore an analysis should depend as little
as possible on the background rates~\cite{mangano}, while a dependence
on the signal rate is unfortunately unavoidable for any new physics
process, which by definition will rely on comparably fewer, rare
events. To show that our analysis fulfills this requirement, and
approximately take into account the unknown NLO QCD effects, we
perform two separate calculations of sensitivity limits:
\begin{enumerate}
\item we assume a uniform $K$-factor of $K=1$ for the $m_{vis}$
  distribution of the background but allow for a normalization
  uncertainty of $\Delta{\cal N}=30\%$ of the SM cross section;
\item we assume a uniform $K$-factor of $K=1.3$ for the $m_{vis}$
  distribution of the background and allow for a normalization
  uncertainty of $\Delta{\cal N}=10\%$ of the SM cross section.
\end{enumerate}
The results from both calculations are then compared and the more
conservative bound is selected. Since the background cross section can
be directly determined from the high $m_{vis}$ region with a
statistical precision of $15\%$ or better for the assumed integrated
luminosities, the bounds we derive should be conservative.

The expression for $\chi^2$ which we use to compute confidence levels
is given by~\cite{babe}
\begin{equation}
\chi^2=\sum_{i=1}^{n_D}\,{(N_i-fN^0_i)^2\over fN_i^0}+(n_D-1)\, ,
\end{equation}
where $n_D$ is the number of bins, $N_i$ is the number of events for a
given $\Delta\lambda_{HHH}=(\lambda-\lambda_{SM})/\lambda_{SM}$, and
$N_i^0$ is the number of events in the SM in the $i$th bin. $f$
reflects the uncertainty in the normalization of the SM cross section
within the allowed range, and is determined by minimizing $\chi^2$:
\begin{equation}
f=\cases{(1+\Delta{\cal N})^{-1} & for $\bar f < (1+\Delta{\cal N})
^{-1}$ \cr
\bar f & for $ (1+\Delta{\cal N})^{-1}<\bar f<1+\Delta{\cal N} $ \cr
1+\Delta{\cal N} & for $\bar f > 1+\Delta{\cal N}$} 
\end{equation}
with
\begin{equation}
\bar f^2=\left\{\sum_{i=1}^{n_D}N_i^0\right\}^{-1}\,\sum_{i=1}^{n_D}{N_i^2
\over N_i^0}~.
\end{equation}
\begin{figure}[t] 
\begin{center}
\includegraphics[width=13cm]{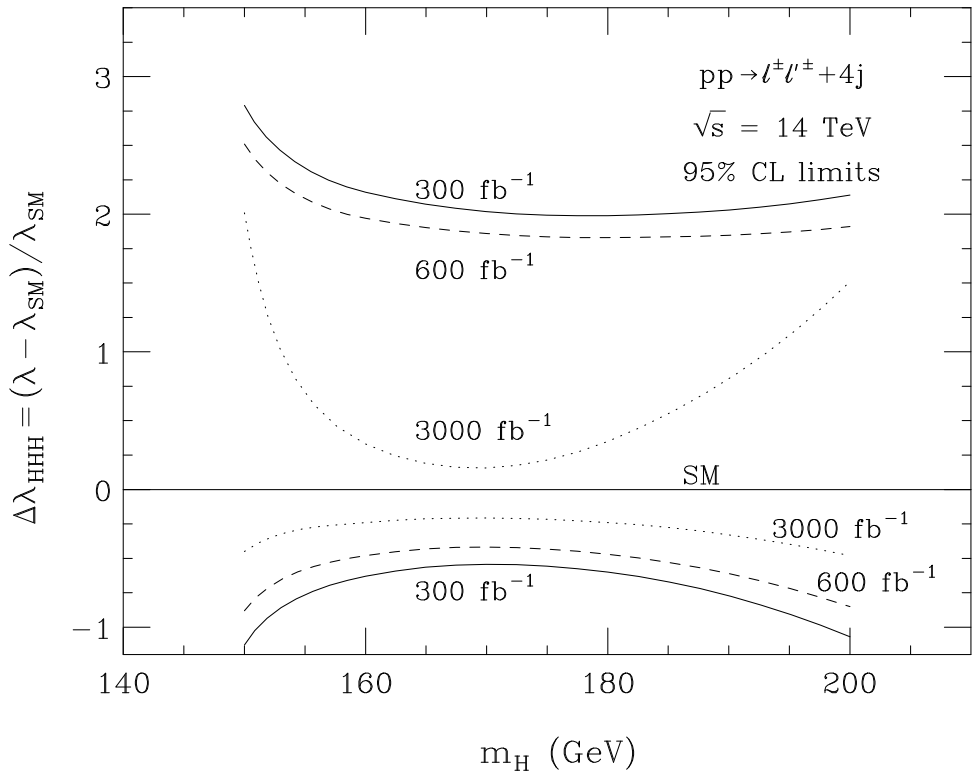} 
\vspace*{2mm}
\caption[]{\label{fig:fig8} 
  Limits achievable at $95\%$ CL for
  $\Delta\lambda_{HHH}=(\lambda-\lambda_{SM})/\lambda_{SM}$ in
  $pp\to\ell^\pm{\ell'}^\pm+4j$ at the LHC. Bounds are shown for
  integrated luminosities of 300~fb$^{-1}$ (solid lines),
  600~fb$^{-1}$ (dashed lines) and 3000~fb$^{-1}$ (dotted lines). The
  allowed region is between the two lines of equal texture.  The Higgs
  boson self-coupling vanishes for $\Delta\lambda_{HHH}=-1$. }
\vspace{-7mm}
\end{center}
\end{figure}
For the LHC, we derive sensitivity limits for integrated luminosities
of 300~fb$^{-1}$, 600~fb$^{-1}$ and 3000~fb$^{-1}$, and Higgs boson
masses in the range $150~{\rm GeV}\leq m_H\leq 200$~GeV. An integrated
luminosity of 300~fb$^{-1}$ (600~fb$^{-1}$) corresponds to 3~years of
running at the LHC design luminosity with one (two) detectors. The
larger value of 3000~fb$^{-1}$ can be achieved in about 3~years of
running at the SLHC with one detector.  Since the cross section for
Higgs boson pair production in the three lepton final state is very
small, we calculate sensitivity bounds only for the same-sign dilepton
channel. Our results are shown in Fig.~\ref{fig:fig8}, which
demonstrates that, for 300~fb$^{-1}$, a vanishing Higgs boson
self-coupling ($\Delta\lambda_{HHH}=-1$) is excluded at the $95\%$ CL
or better, and that $\lambda$ can be determined with a precision of up
to $-60\%$ and $+200\%$.  Doubling the integrated luminosity to
600~fb$^{-1}$ improves the sensitivity by $10-25\%$. For 300~fb$^{-1}$
and 600~fb$^{-1}$, the bounds for positive values of
$\Delta\lambda_{HHH}$ are significantly weaker than those for
$\Delta\lambda_{HHH}<0$, due to the limited number of signal events in
this region of parameter space. At the SLHC, for 3000~fb$^{-1}$, the
Higgs boson self-coupling can be determined with an accuracy of
$20-30\%$ for $160~{\rm GeV}\leq m_H\leq 180$~GeV. The significance of
the SM signal for 300~fb$^{-1}$ (3000~fb$^{-1}$) is slightly more than
$1\,\sigma$ ($3\,\sigma$) for $m_H=150$~GeV and 200~GeV, and about
$2.5\,\sigma$ ($10\,\sigma)$ for Higgs boson masses between 160~GeV
and 180~GeV.  The results shown in Fig.~\ref{fig:fig8} are about
$5-10\%$ weaker than those found in Ref.~\cite{BPR} where only the
dominant $WWWjj$ and $t\bar{t}W$ backgrounds where taken into account
while the effect of all other backgrounds was simulated by multiplying
the combined $WWWjj$ and $t\bar{t}W$ visible invariant mass
distribution by a factor~1.1.

\begin{figure}[h!] 
\begin{center}
\includegraphics[width=12.6cm]{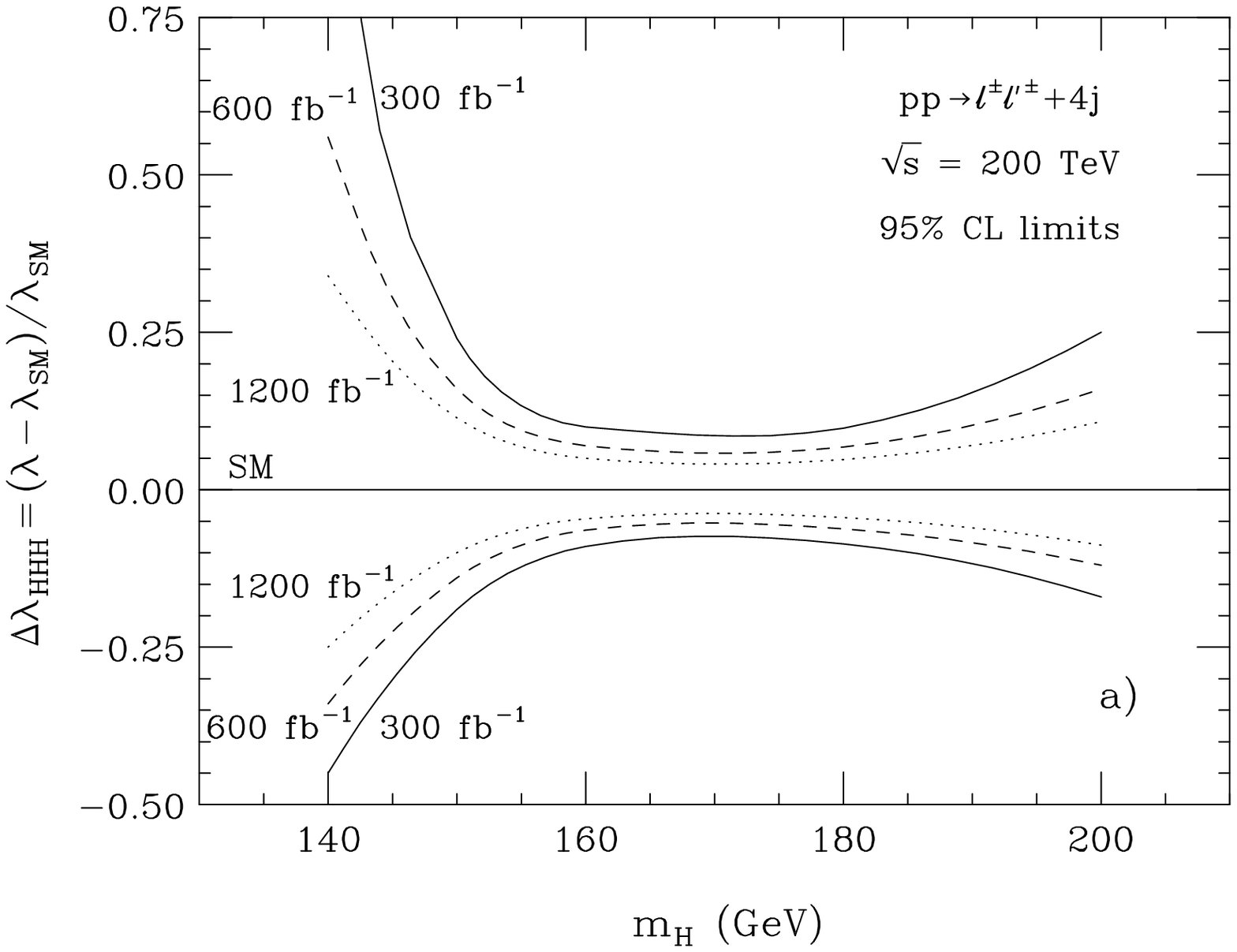} \\[3mm]
\includegraphics[width=12.6cm]{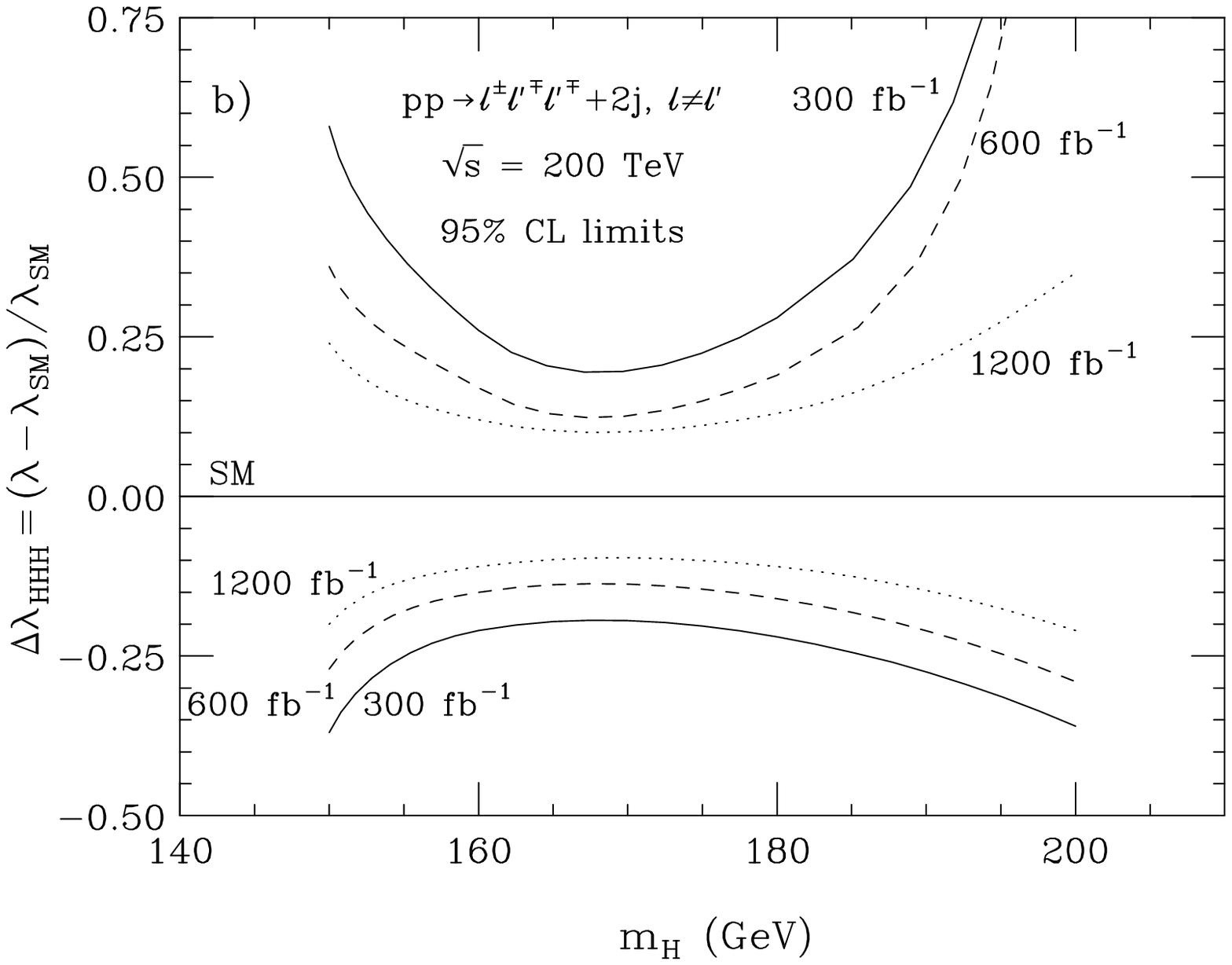}
\vspace*{2mm}
\caption[]{\label{fig:fig9} 
  Limits achievable at $95\%$ CL for
  $\Delta\lambda_{HHH}=(\lambda-\lambda_{SM})/\lambda_{SM}$ for a)
  $pp\to\ell^\pm{\ell'}^\pm+4j$ and b)
  $pp\to\ell^\pm{\ell'}^\mp{\ell'}^\mp+2j$ at the VLHC. Bounds are
  shown for integrated luminosities of 300~fb$^{-1}$ (solid lines),
  600~fb$^{-1}$ (dashed lines) and 1200~fb$^{-1}$ (dotted lines). The
  allowed region is between the two lines of equal texture.  The Higgs
  boson self-coupling vanishes for $\Delta\lambda_{HHH}=-1$.  }
\vspace{-7mm}
\end{center}
\end{figure}

For the VLHC, we calculate bounds for both the
$\ell^\pm{\ell'}^\pm+4j$ and the $\ell^\pm{\ell'}^\mp{\ell'}^\mp+2j$
final states. We assume integrated luminosities of 300~fb$^{-1}$,
600~fb$^{-1}$ and 1200~fb$^{-1}$. For a design luminosity of ${\cal
  L}=2\times 10^{34}~{\rm cm^{-2}\,s^{-1}}$~\cite{vlhc}, the latter
corresponds to 3~years of running with two detectors. The $95\%$ CL
limits which one may hope to achieve at such a machine are shown in
Fig.~\ref{fig:fig9}.

At a $pp$ collider with $\sqrt{s}=200$~TeV and an integrated
luminosity of 300~fb$^{-1}$, the Higgs boson self-coupling can be
measured with a precision of $8-25\%$ at $95\%$ CL for $150~{\rm
  GeV}<m_H<200$~GeV.  For 1200~fb$^{-1}$, the bounds improve to
$4-11\%$. Although the signal to background ratio in the
$\ell^\pm{\ell'}^\mp{\ell'}^\mp+2j$ channel is significantly better,
the sensitivity limits which can be achieved are about a factor~2 to~3
weaker than those obtained for the $\ell^\pm{\ell'}^\pm+4j$ final
state, due to the reduced number of signal events.

Our calculation of sensitivity bounds for $\lambda$ is subject to
several uncertainties which should be addressed in a future more
detailed analysis. In calculating limits, we have ignored the
background from overlapping events and double parton scattering. Their
sizes depend sensitively on the accelerator parameters, in particular
the bunch spacing, and on the ability of the detectors to resolve the
vertices of such events. In addition, these types of background are
difficult to model at the parton level. Our estimates (see
Sec.~\ref{sec:overlap}) indicate that they are small at the LHC, but
may not be negligible at the SLHC or VLHC. A conservative upper limit
of how much the background from overlapping events and double parton
scattering may change the bounds on $\lambda$ is obtained by assuming
that the $m_{vis}$ distribution of such events peaks in the region
where also the signal reaches its maximum. Assuming that this is the
case, and using the results of Table~\ref{tab:two}, one finds that the
SLHC (VLHC) limits weaken by at most $5\%$ ($15\%$).

We also ignored the contributions from $WWZjj$ and $W^+W^-W^+W^-$
production in our calculation. The cross section of these processes is
small compared to that of the dominant background contributions. They
therefore should have a negligible effect on the bounds which can be
obtained. The extremely large number of Feynman diagrams contributing
to $WZjjjj$ and $WWjjjj$ production makes a calculation employing
exact matrix elements currently impractical. To calculate the cross
section for these processes we interfaced the $pp\to WZjj$ and $pp\to
WWjj$ matrix elements with {\sc pythia}. This procedure may well
result in cross sections which differ from the correct result by a
factor $1.5^{\pm 1}$. However, both the $WZjjjj$ and the $WWjjjj$
cross sections are small in the region of $m_{vis}$ where the signal
distribution peaks. Uncertainties in the calculation of their
production cross sections thus should not change the bounds on
$\lambda$ by more than a few per cent. To substantiate this claim, we
have varied the $WZjjjj$ cross section by a factor~$2^{\pm 1}$ and
recomputed the $95\%$ CL limits for $\lambda$. The values obtained
differ from those shown in Figs.~\ref{fig:fig8} and~\ref{fig:fig9} by
at most $5\%$.

Uncertainties in the extraction of sensitivity limits for $\lambda$
also arise from the $t\bar{t}j$ background which we calculated at the
parton level. Since the cross section of the $t\bar{t}j$ background
sensitively depends on the lepton $p_T$ cut and also the lepton
isolation requirement, detector resolution effects may have a
significant effect. Varying the $t\bar{t}j$ cross section by a
factor~$1.5^{\pm 1}$ changes the limits for $\lambda$ by about
$5-10\%$. Finally, QCD corrections are expected to modify the shape of
the $m_{vis}$ distribution for both signal and background. In our
calculation we have approximated the effect of QCD corrections by
uniform $K$-factors which do not take into account this effect. While
an accurate answer how QCD corrections affect the shape of the
$m_{vis}$ distribution requires the calculation of next-to-leading
corrections to signal and background processes, it seems unlikely that
they will change the sensitivity bounds by more than $20\%$.

Our calculation of sensitivity bounds for $\lambda$ has been based on
a simple $\chi^2$ test of the $m_{vis}$ distribution. More powerful
statistical tools, or a neural net analysis, may considerably improve
the limits which can be achieved.

\section{Discussion and Conclusions}
\label{sec:sec5}

A direct experimental investigation of the Higgs potential represents
a conclusive test of the mechanism of electroweak symmetry breaking
and mass generation. After the discovery of an elementary Higgs boson
and the test of its couplings to fermions and gauge bosons,
experimental evidence that the shape of the Higgs potential has the
form required for breaking the electroweak symmetry will complete the
proof that the masses of fermions and weak bosons are generated by
spontaneous symmetry breaking. In order to probe the shape of the
Higgs potential, the Higgs boson self-coupling must be determined.

The Higgs boson self-coupling can be measured in Higgs boson pair
production at lepton or hadron colliders.  In this paper, we presented
a detailed analysis of Higgs boson pair production via gluon fusion
with subsequent decay into four $W$-bosons at the LHC, a luminosity
upgraded LHC (SLHC), and a planned next-generation hadron collider
with a center of mass energy of $\sqrt{s}=200$~TeV (VLHC). We
considered two final states: $\ell^\pm{\ell'}^\pm+4$~jets and
$\ell^\pm{\ell'}^\mp{\ell'}^\mp+2$~jets. To calculate the signal cross
section, exact one-loop matrix elements for finite top quark masses
were used. Final state spin correlations for the $H\to WW\to
4$~fermion decay were fully taken into account, together with finite
$W$ and Higgs boson width effects.

We investigated in detail which processes contribute to the
background, including backgrounds from overlapping events and double
parton scattering. All background cross sections, except those for
$WWjjjj$ and $WZjjjj$ production, were calculated using exact tree
level matrix elements. Contributions to the background from
overlapping events depend on the ability of detectors to resolve
vertex positions, and on machine parameters. We presented a simple
order-of-magnitude estimate of the cross section from overlapping
events which indicates that these should not be a problem at the LHC.
At the SLHC and VLHC, however, the background from overlapping events
could be non-negligible.

At the LHC, the total background cross section is significantly larger
than that of the signal in the $\ell^\pm{\ell'}^\pm+4j$ channel. There
are too few events in the $\ell^\pm{\ell'}^\mp{\ell'}^\mp+2j$ channel
to make it useful.  However, the distribution of the visible invariant
mass of the final state particles, $m_{vis}$, for most of the
processes contributing to the background peaks at a considerably
higher value of $m_{vis}$ than that of the signal, regardless of the
value of $\lambda$. The shape of the $m_{vis}$ distribution can thus
be used as a tool to derive limits on the Higgs boson self-coupling,
$\lambda$.

At the VLHC, we found an improved signal to background ratio for the
$\ell^\pm{\ell'}^\pm+4j$ channel. The
$\ell^\pm{\ell'}^\mp{\ell'}^\mp+2j$ final state has an even more
advantageous signal to background ratio, however, the signal cross
section is significantly smaller than that for $pp\to
HH\to\ell^\pm{\ell'}^\pm+4j$.

In order to determine how well one can hope to measure the Higgs boson
self-coupling at future hadron colliders, we have performed a $\chi^2$
test of the $m_{vis}$ distribution. We found that, at the LHC, with
300~fb$^{-1}$, one will be able to perform a first, albeit not very
precise, measurement of the Higgs boson self-coupling. The
non-vanishing of $\lambda$, however, can be established at $95\%$ CL
or better for $150~{\rm GeV}<m_H<200$~GeV.  {\it This alone is an
  important, non-trivial test of spontaneous symmetry breaking}; the
exact non-zero value of $\lambda$ may vary depending on the way nature
chooses to spontaneously break the electroweak symmetry.  At the SLHC,
for 3000~fb$^{-1}$, a measurement with a precision of up to $20\%$ at
$95\%$ CL is possible; $\lambda$ at the SLHC can be determined with an
accuracy of $10-30\%$ at the $1\,\sigma$ level for Higgs boson masses
between 150 and 200~GeV. Compared with an estimate based on the total
cross section~\cite{blondel}, the fit to the $m_{vis}$ distribution
improves the accuracy of the measurement of Higgs self-coupling by a
factor~1.2 to~2.5.  For the same range of $m_H$, the $95\%$ CL bounds
on $\lambda$ for a 200~TeV $pp$ collider (see Fig.~\ref{fig:fig9})
indicate that deviations of $10\%$ or less from the SM value of
$\lambda$ can be measured at $95\%$ CL if more than 1~ab$^{-1}$ can
been accumulated. Due to the reduced signal rate, limits obtained from
the $\ell^\pm{\ell'}^\mp{\ell'}^\mp+2j$ final state are about a
factor~2 to~3 weaker than those extracted from the
$\ell^\pm{\ell'}^\pm+4j$ channel.

It is interesting to compare the sensitivities which one may hope to
achieve at the LHC, SLHC and VLHC with those obtained for future
$e^+e^-$ linear colliders~\cite{LC_HH3,LC_HH1,LC_HH1a,LC_HH2}. At
TESLA energies, $\sqrt{s}=500-800$~GeV, the Higgs boson self-coupling
can only be determined if $m_H<140$~GeV. For larger values of $m_H$,
the cross section for the dominant Higgs pair production process,
$e^+e^-\to ZHH$, is too small for a useful measurement. For
$m_H=120$~GeV, $\sqrt{s}=500$~GeV, and 1~ab$^{-1}$, one finds that
$\lambda$ can be measured with a precision of $\delta\lambda=\pm 0.20$
for one sigma~\cite{LC_HH1a}.  In contrast, Higgs boson pair
production followed by decays into four $W$ bosons at the LHC and SLHC
offers an opportunity to probe the Higgs boson self-coupling for
masses in the range $150~{\rm GeV}<m_H<200$~GeV. For $m_H<140$~GeV,
where the decay $H\to b\bar{b}$ dominates, the QCD $b\bar{b}b\bar{b}$
background is so large that a measurement of the Higgs boson
self-coupling is hopeless.  LHC and a linear collider operating in the
range of $\sqrt{s}=500-1000$~GeV thus complement each other in their
abilities to determine $\lambda$.

A more direct comparison can be carried out between the VLHC and CLIC, a
proposed multi-TeV $e^+e^-$ linear collider~\cite{CLIC}. For
$m_H=180$~GeV, one finds~\cite{LC_HH3} that, for $e^+e^-$ collisions at
$\sqrt{s}=3$~TeV and an integrated luminosity of 5~ab$^{-1}$, $\lambda$ 
can be determined with a precision of $\delta\lambda=\pm 0.080$
($1\,\sigma$). For the same Higgs boson mass, the Higgs boson
self-coupling can be measured with an accuracy of $\delta\lambda=\pm 0.035$
at a 200~TeV $pp$ collider with 300~fb$^{-1}$. 

Our analysis has been based on leading order parton level
calculations. This introduces uncertainties in our derivation of
sensitivity bounds which we estimated to be of ${\cal O}(20\%)$. 
In order to derive more 
realistic limits for the Higgs boson self-coupling, more
detailed simulations which take into account detector effects, as well
as the effects of higher order QCD corrections are needed.

\acknowledgements
We would like to thank K.~Desch, S.~Dittmaier, I.~Hinchliffe, K.~Jakobs,
F.~Mazzucato, S.~Mrenna, F.~Piccinini, M.~Spira, T.~Stelzer, 
D.~Zeppenfeld and P.M.~Zerwas for useful discussions. 
One of us (U.B.) would like to thank the
Phenomenology Institute of the University of Wisconsin, Madison, and the
Fermilab Theory Group, where part of this work was carried out, for
their generous hospitality and for financial support.
This research was supported in part by the University of Wisconsin
Research Committee with funds granted by the Wisconsin Alumni Research
Foundation, by the U.~S.~Department of Energy under
Contracts No.~DE-FG02-95ER40896 and No.~DE-AC02-76CH03000, and the
National Science Foundation under grants No.~PHY-9970703 and
No.~PHY-0139953. 


\bibliographystyle{plain}

\end{document}